%% file: main.tex
\documentclass[lettersize,journal]{IEEEtran}
\input{util/commands}
\input{util/packages}

\usepackage{amsmath,amsfonts}

\begin{document}

\title{CoFHEE: A Co-processor for Fully Homomorphic Encryption Execution (Extended Version)}

\author{
Mohammed Nabeel\IEEEauthorrefmark{1},~\IEEEmembership{} 
Homer Gamil\IEEEauthorrefmark{1},~\IEEEmembership{} 
Deepraj Soni\IEEEauthorrefmark{2},~\IEEEmembership{} 
Mohammed Ashraf\IEEEauthorrefmark{1},~\IEEEmembership{}
Mizan Abraha Gebremichael\IEEEauthorrefmark{3},~\IEEEmembership{} 
Eduardo Chielle\IEEEauthorrefmark{1},~\IEEEmembership{}
Ramesh Karri\IEEEauthorrefmark{2},~\IEEEmembership{} 
Mihai Sanduleanu\IEEEauthorrefmark{3},~\IEEEmembership{} 
Michail Maniatakos\IEEEauthorrefmark{1} ~\IEEEmembership{}

    \IEEEauthorblockA{\IEEEauthorrefmark{1}Center for Cyber Security, New York University Abu Dhabi, UAE }

    \IEEEauthorblockA{\IEEEauthorrefmark{2}Center for Cyber Security, New York University, USA }

    \IEEEauthorblockA{\IEEEauthorrefmark{3}Khalifa University, Abu Dhabi, UAE }
}


        
        

\markboth{}
{Shell \MakeLowercase{\textit{et al.}}: A Sample Article Using IEEEtran.cls for IEEE Journals}


\maketitle

\input{0_abstract}
\input{1_introduction}
\input{2_preliminaries}
\input{3_method}

\input{3.2_front_end}

\input{3.3_back_end}
\input{4_evaluation}
\input{5_relatedWork}
\input{6_discussion}
\input{7_conclusion}

\bibliographystyle{IEEEtran}
\bibliography{references}

\vfill

\end{document}

%% file: util/commands.tex
\usepackage{mathtools}
\newcommand{\cofhee}{CoFHEE}

\newcommand{\Bigrceil}{{\Bigr\rceil}}
\newcommand{\Biglfloor}{{\Bigl\lfloor}}

\DeclarePairedDelimiter{\floor}{\lfloor}{\rfloor}

\DeclarePairedDelimiter{\Biground}{\Biglfloor}{\Bigrceil}

\newcommand{\um}{\mu m} 
\newcommand{\uW}{\mu W} 


%% file: util/packages.tex
\usepackage{lipsum}
\usepackage{algpseudocode}
\usepackage{algorithm}
\usepackage{mathtools}
\usepackage{amssymb}
\usepackage{amsmath}
\usepackage{multirow}
\usepackage{pifont}
\usepackage{cuted}

\usepackage{float}

\usepackage{soul}
\usepackage{color}
\usepackage{graphics}
\usepackage{subfig}
\usepackage{tablefootnote}
\usepackage{threeparttable}
\usepackage[symbol]{footmisc}

\usepackage{hyperref}
\hypersetup{
    pdfcreator={},   
    pdfproducer={},  
}

%% file: 0_abstract.tex
\begin{abstract}
The migration of computation to the cloud has raised concerns regarding the security and privacy of sensitive data, as their need to be decrypted before processing, renders them susceptible to potential breaches. Fully Homomorphic Encryption (FHE) serves as a countermeasure to this issue by enabling computation to be executed directly on encrypted data. Nevertheless, the execution of FHE is orders of magnitude slower compared to unencrypted computation, thereby impeding its practicality and adoption. Therefore, enhancing the performance of FHE is crucial for its implementation in real-world scenarios. In this study, we elaborate on our endeavors to design, implement, fabricate, and post-silicon validate \cofhee{}, a co-processor for low-level polynomial operations targeting Fully Homomorphic Encryption execution. With a compact design area of $12mm^2$, \cofhee{} features ASIC implementations of fundamental polynomial operations, including polynomial addition and subtraction, Hadamard product, and Number Theoretic Transform, which underlie most higher-level FHE primitives. \cofhee{} is capable of natively supporting polynomial degrees of up to $n = 2^{14}$ with a coefficient size of 128 bits, and has been fabricated and silicon-verified using 55nm CMOS technology. To evaluate it, we conduct performance and power experiments on our chip, and compare it to state-of-the-art software implementations and other ASIC designs.

\end{abstract}


%% file: 1_introduction.tex
\section{Introduction}
The proliferation of cloud services has intensified the user dependency on outsourced computation.
While standard encryption schemes like RSA and AES protect data in-transit and data at-rest, they cannot protect data in-use, since they require data decryption before processing; thus, sensitive data is exposed during computation.
High-profile attacks in cloud services \cite{barron2013cloud} have shown that access control is not sufficient.
The advent of Fully Homomorphic Encryption (FHE) in 2009 \cite{gentry2009thesis} came as a solution for the problem of data in-use as it enables computation to be performed directly in the encrypted domain without the need for decryption.

Although promising from a security and privacy standpoint, FHE is multiple orders of magnitude slower than unencrypted computation, a characteristic that hinders its adoption in the industry.
Much progress has been done in the last decade through improvements in FHE schemes and their software implementation.
However, software-only performance improvement of FHE computation is still not sufficient for the majority of applications. 
In recent years, hardware acceleration of encrypted computation is becoming a means to improve its practicality \cite{dsoni,zhang,soni2023rpu}. 
Several works have been proposed using GPUs \cite{dai2014accelerating} and FPGAs \cite{cousins2012update} for FHE acceleration.
Furthermore, ASIC accelerator protypes demonstrated significant speedups compared to software solutions \cite{gentry2011implementing} \cite{doroz2014accelerating}. To the best of our knowledge, however, there is no fabricated and silicon proven ASIC design supporting FHE computation. 



FHE computation operations can be broadly characterized in two levels: High-level ones, such as key switching and bootstrapping, and low-level, such as polynomial addition and multiplication. All high-level FHE primitives are realized using low-level ones. Recent work~\cite{feldmann2021f1} has shown that a prototype of an ASIC FHE accelerator which includes also high level operations \emph{exceeds} $150mm^2$ (GF12/14) in size, being a shuttle run which can eventually cost many hundreds of thousands of dollars. Therefore, in order to produce a real chip, we need to constrain the design to focus on acceleration of the low-level primitives only. Such a design will eventually serve as a small component in a much bigger design, where the larger design will mostly focus on data movement.

\noindent \textbf{Contributions:} Given a limited design area of $12 mm^2$ available to us given budget constraints, we design an architecture that accelerates the underlying polynomial computation, and it is the base of most high-level FHE operations. This study describes our process for designing, implementing, fabricating, and validating \cofhee{}, the first silicon-proven ASIC co-processor for FHE.
More specifically, our contributions are:
\begin{itemize}
    \item We present the design and implementation of \cofhee{}, a specialized hardware architecture that accelerates low-level operations of FHE
    \item We present a compact low power and wide tuning range All Digital PLL (ADPLL), replacing a traditional PLL to greatly reduce the silicon area occupied
    \item We perform all the physical design preparations to acquire a physical implementation of \cofhee{}, including Floor Planning, Power Planning, Place and Route, and Sign-off Analysis
    \item We fabricate \cofhee{} using 55nm CMOS technology from GlobalFoundries, utilizing a compact design area of $12mm^2$. \cofhee{} is capable of natively supporting polynomial degrees of up to $n = 2^{14}$, with a maximum native coefficient size of 128 bits
    
    \item In addition to the fabrication and implementation of \cofhee{}, we intend to open-source the Register Transfer Level (RTL) code of CoFHEE and its functional units, which can serve as building blocks for future research and development efforts in the design of Fully Homomorphic Encryption (FHE) co-processors
    
\end{itemize}

%% file: 2_preliminaries.tex
\section{Preliminaries}
\label{sec:prelim}

\subsection{Fully Homomorphic Encryption}

Homomorphic encryption is a special type of encryption that enables performing meaningful computations directly on encrypted data.
It can be understood as the functional equivalent of Eq. \ref{eq:he}, where $f(\cdot)$ is a function over plaintexts and $F(\cdot)$ is its equivalent over ciphertexts, $D_k(\cdot)$ and $E_k(\cdot)$ are respectively the decryption and encryption functions for a given key $k$, and $R(\cdot)$ is a function that provides randomness for $E_k(\cdot)$ from ciphertexts $c_a$ and $c_b$.
Although functionally equivalent to Eq. \ref{eq:he}, homomorphic encryption enables equivalent computation on the encrypted domain without the need for any decryption function during computation.

\begin{equation}\label{eq:he}
    F(c_a, c_b) = E_k( f( D_k(c_a), D_k(c_b) ), R(c_a, c_b) )
\end{equation}

While there are several types of homomorphic encryption, Fully Homomorphic Encryption (FHE) possesses at least two orthogonal homomorphic operations allowing any number of arbitrary computations.
FHE was introduced in 2009 \cite{gentry2009thesis} and it is deemed as the "holy grail" of cryptography.
It has seen much progress since its inception with the development of new encryption schemes such as BFV \cite{bfv}, BGV \cite{bgv}, CGGI \cite{cggi}, and CKKS \cite{ckks}.

\subsection{BFV Encryption Scheme}
The Brakerski/Fan-Vercauteren (BFV) scheme is an FHE scheme based on the Ring-Learning With Errors (RLWE) problem \cite{rlwe}.
It works over two polynomial rings, one for the plaintext space and another for the ciphertext space.
The plaintext space is defined over the polynomial ring $\mathcal{P} = \mathbb{Z}_t[x]/(x^n+1)$, while the ciphertext space in defined over $\mathcal{C} = \mathbb{Z}_q[x]/(x^n+1)$, where $n$ is the polynomial degree, and $t$, $q$, and $x^n+1$ are the plaintext, ciphertext, and polynomial moduli, respectively.
BFV is supported by several libraries and frameworks, such as ALCHEMY \cite{alchemy}, Cingulata \cite{cingulata}, E3 \cite{e3}, nGraph-HE2 \cite{ngraph}, Palisade \cite{palisade}, Ramparts \cite{ramparts}, and SEAL \cite{seal}, as well as methods to improve computation speed by combining it with other FHE schemes \cite{bridging}.



\subsection{Ciphertext Multiplication}
\label{sss:pre_ctmul}



The Homomorphic Encryption Security Standard defines the basic primitives of BFV \cite{hestd}.
Out of those primitives, the homomorphic multiplication of ciphertexts (\texttt{EvalMult}) is the slowest operation, and therefore, the main candidate for hardware acceleration.
In order to understand ciphertext multiplication, we first need to understand encryption.
Let $m$ be a plaintext in the plaintext space ($m \in \mathcal{P}$), $k_p = (k_{p1}, k_{p2})$ be an encryption key in the ciphertext space $\mathcal{C}$, and $c = (c_1, c_2) \in \mathcal{C}$ be an encryption of $m$ with key $k_p$;
the encryption function $E(k_p, m) \rightarrow c$ defines a map from $\mathcal{P}$ to $\mathcal{C}$. 
The ciphertext $c$ is composed of two polynomials $c_1$ and $c_2$, which are computed according to Eqs. \ref{eq:encc1} and \ref{eq:encc2}, where $u$ is a random polynomial from the set $\{-1, 0, 1\}$, $e_1$ and $e_2$ are small random polynomials from a discrete Gaussian distribution, and $\Delta = \floor{q/t}$ is a scaling factor.

\begin{align}
    c_1 = k_{p1} \cdot u + e_1 + \Delta m \mod q\label{eq:encc1}\\
    c_2 = k_{p2} \cdot u + e_2 \mod q\label{eq:encc2}
\end{align}

The ciphertext multiplication $c_c = c_a \cdot c_b \Leftrightarrow (c_{c1}, c_{c2}, c_{c3}) = (c_{a1}, c_{a2}) \cdot (c_{b1}, c_{b2})$ is calculated by evaluating the tensor in Eq. \ref{eq:ctmul}.
The term $t/q$ is a scaling factor, while the remaining operations are polynomial multiplications and additions over rings.
The polynomial addition is a simple operation with linear time complexity.
However, a naive implementation of polynomial multiplication has quadratic time complexity.
In addition, there is need  for polynomial reduction after the polynomial multiplication.
More efficient algorithms using the Number Theoretic Transform (NTT) with nega-cyclic convolutions have been proposed \cite{cooley-tukey, gentleman-sande}, reducing the time complexity to $\mathcal{O}(n\log n)$ and avoiding the need for polynomial reduction over the polynomial modulus $x^n+1$.


\begin{equation}\label{eq:ctmul}
\begin{split}
  (c_{c1}, c_{c2}, c_{c3}) = \Big( \Biground{ \frac{t(c_{a1} \cdot c_{b1})}{q} }_q, \\ 
  \Biground{ \frac{t(c_{a1} \cdot c_{b2} + c_{a2} \cdot c_{b1})}{q} }_q, \Biground{ \frac{t(c_{a2} \cdot c_{b2})}{q} }_q \Big)
  \end{split}
\end{equation}

\subsection{Residue Number System}
\label{sss:rns}

As one can see, the coefficient size $\log q$ is usually larger than 64 bits, making computations on 64-bit processors inefficient.
To cope with that, $q$ is usually broken into smaller $q_i$ using the Residue Number System (RNS).
In RNS, a large number is represented by its value modulo several coprime moduli following the Chinese Remainder Theorem.
This effectively breaks each polynomial into several polynomials with smaller coefficients.
In the case of ciphertexts, each pair of polynomials $(c_1, c_2)$ is broken into several pair of polynomials $(c_{1_1}, c_{2_1}), (c_{1_2}, c_{2_2}), (c_{1_3}, c_{2_3}), ...$, called towers.
During ciphertext multiplication, each tower operates independently, and Eq. \ref{eq:ctmul} must be applied to all towers individually.

%% file: 3_method.tex

\section{\cofhee{} Design Flow Overview}
\label{sec:method}

\subsection{Ciphertext Multiplication}
\label{sss:ctmul}

\begin{figure}[t]
    \centering
    \includegraphics[width=\linewidth]{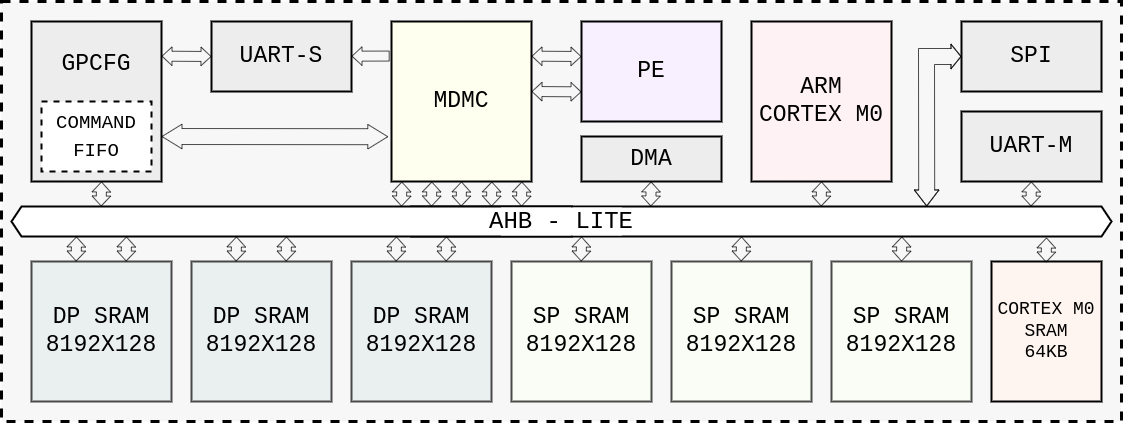}
    \caption{\cofhee{}~Top Level Architecture} 
    \label{fig:TopArch}
\end{figure}

\cofhee{}'s architecture is illustrated in Figure \ref{fig:TopArch}. The most computationally intense low-level FHE operation is the ciphertext multiplication. Consequently, the main objective of the \cofhee{} co-processor is native support for modular operations, which is essential for accelerating FHE cryptosystems. \cofhee{} is designed to natively support polynomial degrees of up to $2^{14}$ (and larger degrees for additional communication costs) in powers of two and modulus size up to 128 bits, while it is optimized to work with a polynomial degree $n = 2^{13}$.
These values offer a trade-off between maximizing performance of FHE applications \cite{kim2015private, pmt, gursoy} and minimizing \cofhee{}'s area.

\cofhee{}'s design is constrained by a limited chip area of $12mm^2$ and the use of $55nm$ CMOS technology from GlobalFoundries. Given the aforementioned limitations and the common encryption parameters used in FHE applications (Section \ref{sss:ctmul}), we have determined that the maximum polynomial degree that can be supported on the chip is $n=2^{13}$, with coefficient sizes of 128 bits. This is the largest coefficient size that can be accommodated within the design area constraints. In the case of $n=2^{14}$, extra communication is required. To achieve this goal, we have developed a system architecture that includes 1 Processing Element (PE), 3 dual-port and 5 single-port SRAMs. This configuration of units allows for full on-chip ciphertext multiplication with an Initiation Interval ($II$) of 1, as the dual-port SRAMs enable the simultaneous fetching and storing of 2 different operands in the same cycle. Moreover, we employed dual-port memory instead of having two single-port memories, because this choice helped in reducing memory layout complexity to load and store multiplier operands/results at the same time, mainly during NTT. It also helped in reducing area (since the area of a single dual-port memory is smaller than that of two single-port memories of the same size) and simplified controller design as both the operands/results required for the PE can be loaded/stored from/to the same memory. As in other works, dual-port memory is not specifically mentioned, we assume they use single-port SRAMs. Hence larger memory layout complexity for NTT.
The majority of the available chip area is occupied by the SRAMs, with the remaining space allocated for the PEs, Multiplier Data Mover and Controller (MDMC), Direct Memory Access controller (DMA), General Purpose Configuration registers (GPCFG), and an ARM Cortex-M0 processor with its own memory. Additionally, the chip operates at a target frequency of $250$ MHz, which is limited by the memory latency, and has two voltage supplies, $3.3$ V for the IO pads and $1.2$ V for the logic core. Table \ref{tab:gpcfg} shows a representative subset of the 35 Configuration Registers in \cofhee{}. The configuration registers in CoFHEE are mapped to the memory range of 0x4002 0000 – 0x4002 FFFF, and the chip's memory base address follows the ARM Cortex M series memory map convention for memory and peripheral addresses. As mentioned before, our on-chip memory includes both dual-port and single-port memory. Dual-port memories are managed by assigning different base addresses to each port, treating them as two distinct address spaces at the bus level. The addition of dual-port memories helped in reducing memory layout complexity especially while loading/storing intermediate results for NTT. It also helped in simpler controller design, especially while performing NTT.

There were other design challenges based on feedback from the physical design, on how to optimally combine multiple banks of memory to form a single logical bank that gives a good floorplan for the chip. As all the other works like F1, Craterlake, etc. are based on simulated results, and not actual fabrication, such optimization challenges are overlooked.


\subsection{Execution \& Operations}
\begin{figure}[t]
    \centering
    \includegraphics[width=\linewidth]{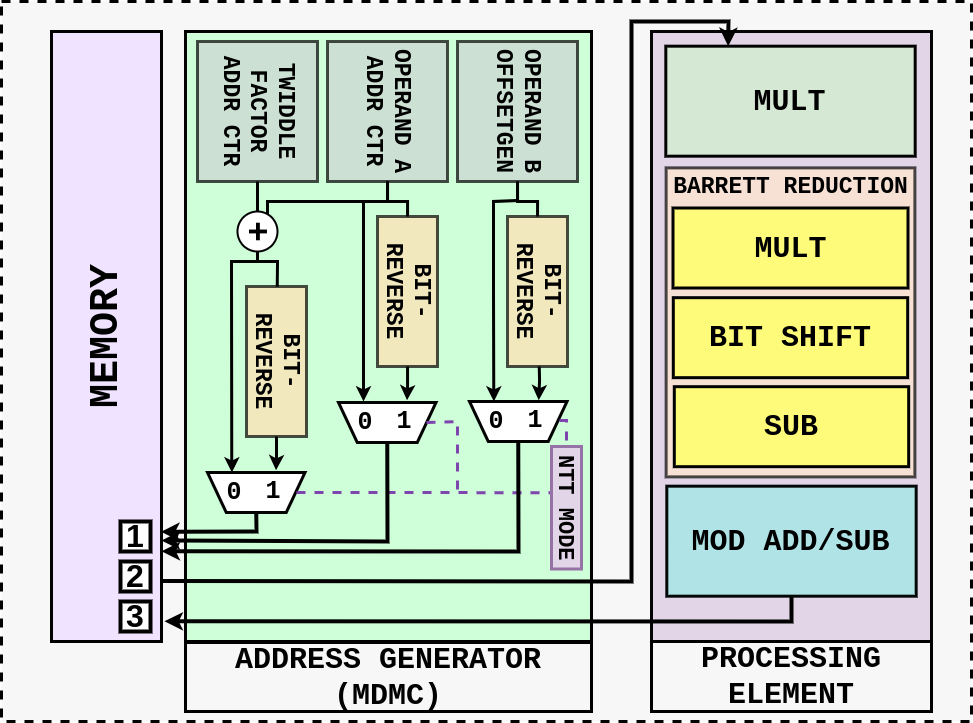}
    \caption{Execution flow of operations} 
    \label{fig:ExecFlow}
\end{figure}




\cofhee{} supports a range of operations that can be broadly classified into two categories: compute and memory operations. Compute operations include the Number Theoretic Transform (NTT), Inverse Number Theoretic Transform (INTT), and a set of pointwise operations such as normal and modular multiplication, modular squaring, modular multiplication by a constant, and modular addition and subtraction. \cofhee{} provides an Instruction Set Architecture (ISA) to execute any of these operations when the relevant polynomials are loaded into the memory. On the other hand, memory operations include the replication or transfer of data from one memory to another. While compute operations run in a sequential manner, memory operations can be run simultaneously, which allows for concurrent execution of both types of operations.
The operations supported by \cofhee{} are summarized in Table \ref{tab:basic_ops}.



\textit{Algorithms}:
Regarding polynomial multiplication, \cofhee{} implements and operates using the Cooley-Tukey NTT algorithm \cite{cooley-tukey}.
\cofhee{} also handles a ciphertext multiplication algorithm. 
Two ciphertexts (polynomial pairs) are passed as input.
The ciphertext polynomials are multiplied following Eq. \ref{eq:ctmul}. 
Each polynomial is used twice in polynomial multiplications.
Nevertheless, we only need one NTT operation per polynomial as they can stay in the NTT domain during the polynomial multiplication and addition.
At the end, we can convert back from the NTT domain using the iNTT operation.
The entire ciphertext multiplication in \cofhee{} utilizes 4 NTTs, 4 Hadamard products, 1 pointwise addition, and 3 iNTTs.

Regarding execution, when executing a command such as NTT, an internal state machine within the MDMC is activated. This state machine is responsible for managing data transfers: specifically, it relocates data from the memory to the processing element (PE) and subsequently transfers the PE's computation results back to the memory.

The state machine also handles the incrementation of addresses for both operands and twiddle factors, although the twiddle factors themselves don't undergo an address increment. The decision regarding how the address is incremented is contingent upon the current stage of the NTT operation. Once a stage is finalized, the MDMC processes the output, subsequently initiating the subsequent stage.
Once the computational operation reaches completion, an interrupt is generated, prompting the command first-in-first-out (FIFO) buffer to issue the succeeding instruction. Figure \ref{fig:ExecFlow} provides an illustration of the execution flow of the proposed method.


\subsection{\cofhee{} API}

\cofhee{} exposes assembly-like instructions for the operations listed in Table \ref{tab:basic_ops}.
Composed operations such as polynomial multiplication or ciphertext multiplication build on top of the atomic ones. 
\cofhee{}'s design is optimized for $n=2^{13}$.
Nevertheless, it supports any $n$, assuming that $n$ is a power of two.
It can perform NTT, polynomial multiplications, and ciphertext multiplications on chip with $II=1$, without requiring back-and-forth communication to the host during the computation for any $n \leq 2^{13}$.
In the case of $n \geq 2^{14}$, \cofhee{} can perform NTT on chip with $II=2$, as single-port memories need to be used for computation. \cofhee{} is also capable of supporting larger polynomials, however, while polynomial multiplications and ciphertext multiplications are possible, they require back-and-forth data movement with the host, as it is not possible to fit all data at once on chip. For larger polynomials the communication costs increase, and the NTT operation becomes more expensive. Nevertheless \cofhee{} can support large polynomials, even if they are not commonly used in FHE applications.

The maximum coefficient size supported by \cofhee{} is 128 bits.
Additions and subtractions take 1 cycle and multiplication takes 4 cycles, independently of the coefficient size.
Coefficients larger than 128 bits must be broken using RNS, similarly to how it is done in software, to fit in \cofhee{}. 

We chose a coefficient size of 128 bits for specific reasons tied to the intricacies and computational burdens associated with Residue Number System (RNS) when breaking down computations into multiple segments. While RNS enables the split of a large modulus polynomial into several smaller modulus towers or limbs, increasing the number of RNS towers also elevates the complexity of computation by introducing additional overheads. Accommodating a larger modulus size allows for fewer RNS towers, resulting in a less expensive execution. Moreover, by utilizing a 128-bit coefficient size, we enable the seamless integration of an efficient key-switching mechanism in future endeavors. It's important to note that key-switching exhibits significantly improved efficiency when operating on 128 bits compared to 32 or 64 bits. This strategic planning ensures a smooth incorporation of key-switching in future developments.

\input{tables/operations}

\subsection{Clock frequency}

\input{tables/gpcfg}

\cofhee{}'s frequency is defined by the critical path of the design as well as the technology node targeted.
The critical path can be either in the computation unit due to a complex data path or it can be in the memory as a consequence of memory read latency.
While the former can be reduced with pipelining, memory read latency is fixed for the technology node and memory size used.
As expected, \cofhee{}'s critical path is in the memory read for loading data from the memory to the computation unit.
For the technology node targeted, i.e., Globalfoundries (GF) 55nm low power enhanced (LPE) process, the memory read path is around $4ns$, which translates to a clock frequency of 250MHz.

\subsection{Processing Element}
\cofhee{} comprises a singular modular multiplier, along with modular adder and subtractor units. Furthermore, the processing element encompasses a series of multiplexers (MUXs) designed to direct operands and intermediate data, depending on whether the operation involves multiplication, addition, or subtraction within the utilized butterfly structure. This enables the processing element to function as a versatile unit for NTT and iNTT operations, as well as element-wise additions, subtractions, and multiplications.

Surrounding the processing element, we have implemented a wrapping logic that orchestrates the movement of data from memory to the processing element, tailored to the specific operation at hand, whether it's NTT, iNTT, or element-wise operations. In the context of NTT and iNTT, the processing element essentially serves as a singular radix-2 butterfly NTT unit.

\cofhee{}'s PE supports modular multiplication, modular addition, and modular subtraction.
Each operation has an $II=1$.
Modular addition and subtraction have a latency of one clock cycle, while modular multiplication completes in five clock cycles.
In contrast to the state-of-the-art that uses Montgomery multipliers, \cofhee{} employs a Barrett multiplier to implement multiplication and modular reduction due to its pipelining capabilities \cite{modmulalgocomp}. The pipeline depth is chosen such that the critical path matches the memory read latency. 
Our PE operates in four distinct modes defined by the mode selection input.
The operations modes are: (1) Modular Multiplication, (2) Modular Addition, (3) Modular Subtraction, and (4) Butterfly operation.
The butterfly operation is an atomic computational unit of NTT.
It is executed by performing modular multiplication followed by modular addition and modular subtraction.
We implement the butterfly operation using radix 2, and acquire a PE that occupies $6\%$ of the design area.
\subsection{NTT considerations}
For a $2\times$ performance improvement of NTT, the input and output polynomials are stored in dual-port memories.
In addition, there is a third dual-port memory to store the next polynomial to go under NTT.
If the next polynomial is in a single-port memory, DMA is used to transfer polynomial coefficients to the available dual-port memory while NTT operates on the current polynomial.
This is possible since the bus architecture allows the MDMC, DMA, and ARM CM0 to access memories in parallel.
Once NTT completes, the dual-port memories switch tasks; the one containing the next polynomial participates in the following NTT, while the memory containing the output uses DMA to offload the result to a single-port memory and load the next polynomial for NTT.
This process happens transparently in the background without performance degradation due to data movement.

\subsection{Internal Data Flow}
\subsubsection{Interconnect}
We implement a parameterized Advanced High-Performance Bus (AHB) lite interconnect \cite{noauthor_amba_2001} as it supports the required bus transfer operations and has low area utilization. We opted for a lightweight AHB-Lite bus which has less area, latency, and signal counts (helps in less congestion at the backend). F1 \cite{feldmann2021f1}, on the contrary, uses three 16x16 512-byte cross bars, each having an area of $3.33 mm^2$ in a 12nm technology node. CoFHEE’s bus area is $0.07 mm^2$ in 55nm technology node ($0.008192 mm^2$ after applying the 55-to-14nm scaling factor)  and it is a 10x11-152 byte crossbar.
\cofhee{} requires single memory transfers as well as burst memory operations, with the data sizes ranging from 32 to 128 bits. 
Our configuration registers map to the 0x4002\_0000 -- 0x4002\_FFFF memory range.
In our design, the memory base address follows the ARM Cortex M series memory map convention for memory and peripheral addresses.

\subsubsection{Multiplier Data Mover and Controller (MDMC)}

The sequence of commands to be executed is stored in the command FIFO, which decodes the command and triggers the MDMC for the requested operation. The FIFO also provides the memory base addresses of input operands and output result.
When running computation-based operations, the MDMC fetches data from memory and forwards to the PE on every clock cycle until the operation is completed.
Once the data is processed, the MDMC stores the data back to the output memory.
In order to achieve an $II=1$ for each butterfly operation, we store the input and output polynomials in dual-port memories.
Thus, we can fetch two polynomial coefficients from one memory (dual-port) and a twiddle factor from another (single-port) in a single clock cycle.
After computing the butterfly operation, the MDMC stores the two outputs in the output memory (dual-port).
Once an NTT stage completes, the output memory acts as input memory and vice-versa, until the NTT/iNTT is finalized.

\subsection{External Interfaces}
\cofhee{} provides SPI and UART interfaces for external host communication.
These interfaces are used for loading polynomials, triggering the required operation and reading back the result.
SPI and UART are chosen mainly for their simplicity in implementation as well as the ability to communicate from an external PC.
One can always replace these interfaces with faster ones such as PCI-express or HSIC (High-Speed Inter-Chip).
Other IOs in \cofhee{} are clock, reset, voltage supplies, bias voltages, bias currents, PLL controls, and debugging IOs.





\subsection{Execution modes}
There are three ways to execute the basic operations supported by \cofhee{} (Table \ref{tab:basic_ops}). The simplest option is for the external host to directly trigger the MDMC to perform the commands through a configuration register write.
This mode is slow as there are delays imposed by the communication interface when writing to the configuration register.
A second possibility is to use the command FIFO, where the external host preloads the sequence of commands to be executed and waits for an interrupt issued by \cofhee{} signaling that the queue is empty.
As soon as the first command is written to the queue, the command FIFO sends it to the MDMC.
The MDMC starts the operation and once it completes, it sends a signal to the command FIFO.
The process repeats until all commands have been executed.
The command FIFO guarantees the execution of a single command at a time in a predefined order.
Thus, it requires less control logic and avoids complicated  out-of-order executions.
In addition, the command FIFO gives the flexibility to execute different complex operations as combinations of basic instructions. 
We define the length of the queue to be 32 commands, as it is more than sufficient for our target applications. 
The host can continuously load more commands while the queue is not full. Lastly, for a faster and flexible sequencing and execution of commands, we introduce a third mode, which utilizes a 32-bit ARM Cortex M0 (CM0) along with a dedicated instruction memory.
In this mode, the processor is used instead of the command FIFO.
One can write complex subroutines and sequence of operations in embedded C, then compile and preload it in CM0's instruction memory for later execution.

\subsection{Pre-silicon verification}

We verified the functionality of our RTL design using both simulation and FPGA-based validation.
The simulation was performed using Synopsys VCS at the top-level.
A python script is used to calculate the modulus following the equation $q = 2 k \cdot n + 1$, where $k \ge 1$ is an arbitrary constant.
In addition, the script finds twiddle factors, generate random input polynomial coefficients, and calculate expected results.
We use random coefficient values modulo $q$ for our test polynomials since the 128-bit operand range cannot be exhaustively tested.
These values are then ported to the verilog testbench, where they are loaded and the result from the design is compared against the expected result.
Moreover, for our FPGA design, we implemented a scaled-down version of \cofhee{}, as $n=2^{13}$ is incompatible with the available resources of our FPGAs.
Specifically, the maximum polynomial degree that could be supported on a Digilent Nexys 4 is $n=2^{12}$ running at 10 MHz.

\subsection{Synthesis}

\cofhee{}'s RTL code is synthesized using a $55nm$ standard cell library from GlobalFoundries and a clock constraint of 250MHz.
SPI IO timing is constrained to meet 50MHz of interface speed.
There is no specific IO timing constraint set for UART as it is an asynchronous interface.
Following standard practices, the standard cell library used for synthesis was the one characterized for the worst voltage (1.08V), temperature (125C), resistance, and capacitance.
Synthesizing with such a library ensures that we can achieve the target frequency in various operating conditions.
For synthesis, we use the Synopsys Design Compiler (DC).
In Table \ref{tab:area}, we present the post synthesis area and timing of the major \cofhee{} blocks.
Other than memory, the largest design is the PE, followed by the AHB and configuration registers.
Post-synthesis, we observe many blocks with critical timing path much higher than the target clock period of $4ns$. This is because synthesis setup uses restricted standard cell library with no access to faster library cells such as the ones with lower threshold voltage $V_t$.
As these paths are long combinational paths, they easily meet timing in the backend, where all the standard cell libraries are used.
In this condition, paths starting from memory become the critical path due to memory read latency (around $3.1ns$). 
We verified the functionality of our RTL design using both simulation and FPGA-based validation.

%% file: tables/operations.tex
\begin{table}[t!]
\centering
\resizebox{\columnwidth}{!}{%
    \begin{tabular}{||l|cccccccccc|l||}
        \hline
        \multirow{2}*{\textbf{Command}} & \multicolumn{10}{c|}{\textbf{Inputs}} & \multirow{2}*{\textbf{Operations}}  \\
        
        {}      & $n$       & $[\vec{x}]$ & $[\vec{y}]$ & $[\vec{\omega}]$ & $q$    & $n^{-1}$  & $[\vec{t}]$ & $\delta$  & $\Rsh$    & $\rightarrowtail$ & {} \\ \hline \hline
        
        NTT     & $\bullet$ & $\bullet$ & {}        & $\bullet$ & $\bullet$ & {}        & $\bullet$ & {}        & {}        & {}        & Performs NTT on $\vec{x}$. \\ \hline
        
        \multirow{2}*{iNTT}    & \multirow{2}*{$\bullet$} & \multirow{2}*{$\bullet$} & {} & \multirow{2}*{$\bullet$}
        & \multirow{2}*{$\bullet$} & \multirow{2}*{$\bullet$} & \multirow{2}*{$\bullet$} & {} & {} & {} & Performs inverse NTT \\
        {} & {} & {} & {} & {} & {} & {} & {} & {} & {} & {} & on $\vec{x}$. \\ \hline
        
        \multirow{2}*{PMODADD} & \multirow{2}*{$\bullet$} & \multirow{2}*{$\bullet$} & \multirow{2}*{$\bullet$}
        & {} & \multirow{2}*{$\bullet$} & {} & {} & {} & {} & \multirow{2}*{$\bullet$} & Pointwise modular \\
        {} & {} & {} & {} & {} & {} & {} & {} & {} & {} & {} & addition of $\vec{x}$ and $\vec{y}$. \\ \hline
        
        \multirow{2}*{PMODMUL} & \multirow{2}*{$\bullet$} & \multirow{2}*{$\bullet$} & \multirow{2}*{$\bullet$}
        & {} & \multirow{2}*{$\bullet$} & {} & {} & {} & {} & \multirow{2}*{$\bullet$} & Pointwise mod. multi- \\
        {} & {} & {} & & {} {} & {} & {} & {} & {} & {} & {} & plication of $\vec{x}$ and $\vec{y}$. \\ \hline
        
        \multirow{2}*{PMODSQR} & \multirow{2}*{$\bullet$} & \multirow{2}*{$\bullet$} & {} 
        & {} & \multirow{2}*{$\bullet$} & {} & {} & {} & {} & \multirow{2}*{$\bullet$} & Pointwise modular \\
        {} & {} & {} & & {} {} & {} & {} & {} & {} & {} & {} & squaring of $\vec{x}$. \\ \hline
        
        \multirow{2}*{PMODSUB} & \multirow{2}*{$\bullet$} & \multirow{2}*{$\bullet$} & \multirow{2}*{$\bullet$}
        & {} & \multirow{2}*{$\bullet$} & {} & {} & {} & {} & \multirow{2}*{$\bullet$} & Pointwise mod. sub- \\
        {} & {} & {} & {} & {} & {} & {} & {} & {} & {} & {} & traction of $\vec{x}$ and $\vec{y}$. \\ \hline
        
        \multirow{2}*{CMODMUL} & \multirow{2}*{$\bullet$} & \multirow{2}*{$\bullet$} & {}
        & {} & \multirow{2}*{$\bullet$} & {} & {} & {} & {} & \multirow{2}*{$\bullet$} & Mod. multiplication \\
        {} & {} & {} & & {} {} & {} & {} & {} & {} & {} & {} & of $\vec{x}$ by a constant. \\ \hline
        
        \multirow{2}*{PMUL}    & \multirow{2}*{$\bullet$} & \multirow{2}*{$\bullet$} & \multirow{2}*{$\bullet$}
        & {} & {} & {} & {} & {} & {} & \multirow{2}*{$\bullet$} & Pointwise multiplica- \\
        {} & {} & {} & {} & {} & {} & {} & {} & {} & {} & {} & tion of $\vec{x}$ and $\vec{y}$. \\ \hline
        
        \multirow{2}*{MEMCPY} & {} & {} & {} & {} & {} & {} & {} & \multirow{2}*{$\bullet$}
        & \multirow{2}*{$\bullet$} & \multirow{2}*{$\bullet$} & Memory-to-memory \\
        {} & {} & {} & {} & {} & {} & {} & {} & {} & {} & {} & data transfer. \\ \hline
        
        \multirow{2}*{MEMCPYR} & {} & {} & {} & {} & {} & {} & {} & \multirow{2}*{$\bullet$}
        & \multirow{2}*{$\bullet$} & \multirow{2}*{$\bullet$} & Memory data transfer \\
        {} & {} & {} & {} & {} & {} & {} & {} & {} & {} & {} & in bit-reverse. \\ \hline
    \end{tabular}
    }
    \vspace{1em}
    \caption{\cofhee's operations. $[\cdot]$: memory address function, $n$: polynomial degree, $\vec{x}$ and $\vec{y}$: polynomials, $\vec{\omega}$: twiddle factors, $q$: modulus, $n^{-1}$: inverse of $n$, $\vec{t}$: temporary values, $\delta$: length (in words), $\Rsh$: source address, $\rightarrowtail$: output/destination address}
    \vspace{-1em}
    \label{tab:basic_ops}
\end{table}

%% file: tables/gpcfg.tex
\begin{table}[t!]
    \centering
    \resizebox{\columnwidth}{!}{%
    
    \begin{tabular}{||l|l|c||}
         \hline
         \multirow{2}{*}{\parbox{2.7cm}{{\bf Register Name}}}                             &
         \multirow{2}{*}{\parbox{4.1cm}{{\bf Description}}}                          &
         \multirow{2}{*}{\parbox{0.7cm}{\centering {\bf Bit Size}}}                                   \\
         {} & {} & {} \\
         \hline \hline
         UARTMTX\_PAD\_CTL     &   IO pad control for primary UART TX    &    32 \\ \hline
         UARTMRX\_PAD\_CTL     &   IO pad control for primary UART RX    &    32 \\ \hline
         UARTSTX\_PAD\_CTL     &   IO pad control for secondary UART TX  &    32 \\ \hline
         SPIMOSI\_PAD\_CTL     &   SPI data in pad control      & 32 \\ \hline
         SPIMISO\_PAD\_CTL     &   SPI data out pad control     & 32 \\ \hline
         SPICLK\_PAD\_CTL      &   SPI clock pad control        & 32 \\ \hline
         SPICSN\_PAD\_CTL      &   SPI chip select pad control  & 32 \\ \hline
         HOSTIRQ\_PAD\_CTL     &   IO pad control for Host Interrupt  &    32 \\ \hline
         UARTM\_BAUD\_CTL      &   Baud control for primary UART  &    32 \\ \hline
         UARTS\_BAUD\_CTL      &   Baud control for secondary UART  &    32 \\ \hline
         UARTM\_CTL            &   Primary UART control   &    32 \\ \hline
         UARTS\_CTL            &   Secondary UART control  &    32 \\ \hline
         SIGNATURE              &   Stores Chip ID  &    32 \\ \hline
         Q                      &   Modulus $q$                                  &    128 \\ \hline
         N                      &   Polynomial degree $n$                        &    128 \\ \hline
         INV\_POLYDEG           &   $n^{-1} \ \mod q$                            &    128 \\ \hline
         BARRETTCTL1            &   $barrettk = 2 \cdot \log n$                  &    32  \\ \hline
         BARRETTCTL2            &   barrett constant = $2^{barrettk}/q$          &    160 \\ \hline
         FHECTL1                &   Command FIFO select and $n$    &    32 \\ \hline
         FHECTL2                &   Trigger bits for different commands          &    32 \\ \hline
         FHECTL3                &   Select or bypass PLL clock          &    32 \\ \hline
         PLLCTL                &   Control bits  required for the PLL          &    32 \\ \hline
         COMMANDFIFO            &   Trigger bits for different commands          &    32 \\ \hline
         DBG\_REG               &   Debug register   &    32 \\ \hline
    \end{tabular}
    }
    \caption{Subset of \cofhee{} Configuration Registers.} \label{tab:gpcfg}
\end{table}

%% file: 3.2_front_end.tex
\section{Functional Units}
\label{sec:fun_units}

\subsection{Modular Multiplier Unit}
Modular multiplication involves normal multiplication followed by a modular reduction of the product. There are few popular ways to implement modular multipliers; interleaved multiplier, Barrett multiplier, and Montgomery multiplier \cite{cryptohandbook}. 
Barrett is selected for our implementation as there is no need to transform the arguments, as required for Montgomery \cite{modmulalgocomp}. 
\subsection{NTT Unit}
\label{ss:ntt}

The Number Theory Transform (NTT) is the generalized Discrete Fourier Transform (DFT).
It is an integral part of FHE algorithms for accelerating polynomial multiplications over finite fields.
Without NTT, polynomial multiplications have quadratic complexity $\mathcal{O}(n^2)$.
NTT converts polynomials to a domain that reduces the complexity of polynomial multiplications to linear $\mathcal{O}(n)$.
There are several algorithms for NTT \cite{cooley-tukey, gentleman-sande}; \cofhee{} implements the Cooley-Tukey algorithm \cite{cooley-tukey} as shown in Algorithm \ref{alg:ntt}.
Each of the $\log n$ stages of NTT computes $n/2$ butterfly operations.
As discussed earleir, the butterfly operation is an atomic computational unit of the NTT. The operation is comprised of the primitive operations of modular multiplication, modular addition, and finally, modular subtraction. 
\subsection{Polynomial Multiplication Unit}
Polynomial multiplication over rings is the main operation in ciphertext multiplication for RLWE-based FHE.
This operation can be divided into two parts: (1) Multiplication and (2) Reduction.
1) For the first part, we use NTT to reduce the complexity of the operation from $\mathcal{O}(n^2)$ to $\mathcal{O}(n\log n)$, as discussed in Section \ref{ss:ntt}.
2) The second part reduces the polynomial over the polynomial modulus.
This operation is avoided using Negative Wrapped Convolution (NWC).
NWC requires the $2n^{th}$ primitive root of unity and reduces the polynomial degree to $n-1$ over the polynomial $x^n+1$.
Algorithm \ref{alg:polynomial_mult} describes the polynomial multiplication operation supported by \cofhee{}.

\subsection{Ciphertext Multiplication}
The ciphertext multiplication is the slowest operation in FHE computation.
We discuss it in detail in Section \ref{sss:ctmul}.
Algorithm \ref{alg:ctmul} describes this operation as handled by \cofhee{}.
Two ciphertexts (polynomial pairs) are passed as input.
The ciphertext polynomials are multiplied following Eq. \ref{eq:ctmul}. 
Each polynomial is used twice in polynomial multiplications.
Nevertheless, we only need one NTT operation per polynomial as they can stay in the NTT domain during the polynomial multiplication and addition.
At the end, we can convert back from the NTT domain using the iNTT operation.
The entire ciphertext multiplication in \cofhee{} utilizes 4 NTT operations, 4 Hadamard products, 1 pointwise addition, and 3 iNTT operations.


\begin{figure}[ht!]
    \begin{minipage}[t][][t]{0.46\textwidth}
        \centering
        \input{algo/ntt}
        \input{algo/polymul}
    \end{minipage}
    \hspace{0.2cm}
    \begin{minipage}[t][][t]{0.46\textwidth}
        \centering
        \input{algo/ctmul}
    \end{minipage}
\end{figure}

%% file: algo/ntt.tex
\begin{algorithm}[H]
\small
\caption{Number Theory Transform}
{\bf Input:} polynomial ($\vec{x}$)\\
{\bf Input:} roots of unity ($\vec{\omega}$)\\
{\bf Input:} modulus ($q$)\\
{\bf Output:} polynomial ($\vec{x}$)
\vspace{0.3em}
\begin{algorithmic}[1]
  \State{idx = 0}
  \For{$i\gets n/2$; $i >= 2$;$i = i >> 1 $}
    \For{$j\gets 0$; $j < n/2$;$j = j + i $}
      \State{twiddle = $\vec{\omega}$[idx++]}   
      \For{$k\gets j$; $k < j + i$; $k = k + 1 $}
        \State{m = twiddle\ * $\vec{x}$[k+i]}
        \State{m = reduce$_q$(m)}
        \State{$\vec{x}$[k+i] = reduce$_q$($\vec{x}$[k] - m)}
        \State{$\vec{x}$[k] = reduce$_q$($\vec{x}$[k] + m)}
      \EndFor
    \EndFor
  \EndFor
\end{algorithmic}
\label{alg:ntt}
\end{algorithm}

%% file: algo/polymul.tex
\begin{algorithm}[H]
\small
\caption{Polynomial Multiplication}
\bf{Input:} $A(x)$, $B(x) \in \mathbb{Z}_q[x]/x^n+1$\\
\bf{Input:} $n^{th} \ \text{roots of unity} \ \vec{\omega} \in \mathbb{Z}_q$ \\
\bf{Input:} $2n^{th} \ \text{roots of unity} \ \vec{\psi} \in \mathbb{Z}_q$ \\
\bf{Output:} $Y(x) \in \mathbb{Z}_q[x]/x^n+1$ \\
$\centerdot \quad Y(x) = A(x) \times B(x)$
\vspace{0.3em}
\begin{algorithmic}[1]
\State{ $A'(x) = \text{NTT}((A(x) \ \bullet \ \vec{\psi}),\  \vec{\omega})$ }
\State{ $B'(x) = \text{NTT}((B(x) \ \bullet \ \vec{\psi}),\  \vec{\omega})$ }
\State{ $Y'(x) = A'(x) \ \bullet \ B'(x)$ }
\State{ $Y(x) = \text{iNTT}((Y'(x) \ \bullet \ {\vec{\psi}}^{-1}),\ \vec{\omega}$) } \label{lst:line:intt}

\State {\bf return} $Y(x)$
\end{algorithmic}
\label{alg:polynomial_mult}
\end{algorithm}



%% file: algo/ctmul.tex

\begin{algorithm}[H]
\small
\caption{Ciphertext Multiplication}
\bf{Input:} $(A_0(x), A_1(x)) \in \mathbb{Z}_q[x]/x^n+1$ \\
\bf{Input:} $(B_0(x), B_1(x)) \in \mathbb{Z}_q[x]/x^n+1$ \\
\bf{Input:} $n^{th} \ \text{roots of unity} \ \vec{\omega} \in \mathbb{Z}_q$ \\
\bf{Input:} $2n^{th} \ \text{roots of unity} \ \vec{\psi} \in \mathbb{Z}_q$ \\
\bf{Output:} $Y_i(x) \in \mathbb{Z}_q[x]/x^n+1 \  \forall i \in [0, 2]$ \\
$\centerdot \quad Y_0(x) = A_0(x) \times B_0(x)$ \\
$\centerdot \quad Y_1(x) = A_0(x) \times B_1(x) + A_1(x) \times B_0(x)$ \\
$\centerdot \quad Y_2(x) = A_1(x) \times B_1(x)$
\vspace{0.3em}
\begin{algorithmic}[1]
\State{ $B_0'(x) = \text{NTT}((B_0(x) \ \bullet \ \vec{\psi}),\  \vec{\omega})$ }
\State{ $A_0'(x) = \text{NTT}((A_0(x) \ \bullet \ \vec{\psi}),\  \vec{\omega})$ }
\State{ $Y_0'(x) = A_0'(x) \ \bullet \ B_0'(x)$ }
\State{ $Y_0(x) = \text{iNTT}((Y_0'(x) \ \bullet \ {\vec{\psi}}^{-1}),\ \vec{\omega}$) }
\State{ $B_1'(x) = \text{NTT}((B_1(x) \ \bullet \ \vec{\psi}),\  \vec{\omega})$ }
\State{ $Y_{01}'(x) = A_0'(x) \ \bullet \ B_1'(x)$ }
\State{ $A_1'(x) = \text{NTT}((A_1(x) \ \bullet \ \vec{\psi}),\  \vec{\omega})$ }
\State{ $Y_2'(x) = A_1'(x) \ \bullet \ B_1'(x)$ }
\State{ $Y_2(x) = \text{iNTT}((Y_2'(x) \ \bullet \ {\vec{\psi}}^{-1}),\ \vec{\omega}$) }
\State{ $Y_{10}'(x) = A_1'(x) \ \bullet \ B_0'(x)$ }
\State{ $Y_1'(x) = Y_{01}'(x) + Y_{10}'(x)$ }
\State{ $Y_1(x) = \text{iNTT}((Y_1'(x) \ \bullet \ {\vec{\psi}}^{-1}),\ \vec{\omega}$) }
\State {\bf return} $(Y_0(x),Y_1(x),Y_2(x))$

\end{algorithmic}
\label{alg:ctmul}
\end{algorithm}


%% file: 3.3_back_end.tex
\section{Physical Design}
\label{sec:backend}

\input{tables/pnr}

\input{figs/backend}

\begin{figure}[t]
    \begin{minipage}[t][][t]{0.5\textwidth}
        \input{tables/layout}
    \end{minipage}
\end{figure}

\begin{figure*}[t]
    \begin{minipage}[t][][t]{0.33\textwidth}
        \input{tables/performance}
    \end{minipage}
    \hspace{0.5cm}
        \begin{minipage}[t][][t]{0.33\textwidth}
        \input{tables/eda_tools}
    \end{minipage}
    \hspace{0cm}
        \begin{minipage}[t][][t]{0.33\textwidth}
        \input{tables/redundant_vias}
    \end{minipage}
\end{figure*}

This section covers \cofhee{}'s physical design aspects.
Table \ref{tab:stages_tools} summarizes the stages and EDA tools utilized. \cofhee{}'s total die area, including the seal ring, is $15 mm^2$. We also note that we have implemented the use of inline pads placed on all four sides of the chip.
Table \ref{tab:design_features} provides information about the signal and power/ground pads, obtained from Synopsys Design Compiler. Table \ref{tab:layout} shows the physical parameters after place and route multiple iterations.

\subsection{Floor Planning}

The die dimensions are $3660 \um \times 3842 \um$.
Fig. \ref{fig:floorplan} shows the design's floorplan, which focused on utilizing the maximum available silicon area. The overlapping regions within the corner pads are empty, thus they do not cause any DRC/LVS concerns. At the same time, they maximize the available area for logic gates or memories placement. The Phase-Locked Loop (PLL) is placed  at the upper right corner of the chip. IO pads dedicated for the PLL are also placed in the same corner.
Overall, there are 68 memory instances, out of which 48 (16x2096) are dual-port, and 16 (32x8192) plus 4 (32x4096) are single-port.
There are 47 digital IO pads including power pads.
There are two pads for VDD (core power supply), VSS (core ground), and IO power/ground (DVDD/DVSS).
Furthermore, in this stage we place edge cells at the boundaries, and well tap cells throughout the layout in a staggered fashion (according to the foundry specifications) to avoid latchup violations.

\subsection{Power Planning}
In this section we discuss our efforts regarding power planning. Our power network consists of four pairs of VDD/VSS rings around the core region.
Out of the eight metal layers, the top two (BA and BB)
are used for rings.
Furthermore, all the power and ground pads are connected to the aforementioned rings through multiple connections, as shown in Fig. \ref{fig:powerpad}.
Multiple power and ground straps in layers BA, BB, M5, and M4 run over the entire core region at different pitches of $30 \um$ (BA/BB), and $50 \um$ (M4/M5).
M1 rails are connected directly to M4 straps through stacked vias.
Fig. \ref{fig:powernetwork} provides a high-level illustration of our power/ground network, which includes power rings, straps, standard cell power rails, and connections to pad and memory pins.
The power straps are connected to each other in the power network until M4, which delivers power to the standard cell rails through stacked vias.
The connection, in Fig. \ref{fig:powerconnection}, is done, so that power straps in M2 are avoided, else there can be standard cell pin access issues.
Optimal pitches were selected after multiple iterations of analyzing the routing congestion, IR drop and effective resistance.
As it is a memory dominant design, delivering power in all the channels between the memories was another challenge we faced.
The flow was modified to ensure that every such channel is delivered power and ground sufficiently, and that the memory power ground pins run on top of them, on layer M4.
The connections to the memories power/ground pins are made by dropping vias from M5 straps which run over them.

\subsection{Place and Route}
The design is then taken through place and route.
Placement optimization is done in one slow and one typical corner.
No-buffering blockages are added in the narrow channels between memories to
prevent the placement of logic cells in the channels, as placing them in the wrong location can result in either timing, and congestion issues, or affect the clock tree synthesis.
Fig. \ref{fig:placement} shows the module placement.
Post placement, the clock tree synthesis is done using standard rules like double width and double spacing for the clock nets. In addition, it is verified that clock buffers and inverters used are of optimal drive strength, so that they don't contribute much to IR drop.
Fig. \ref{fig:clocknetwork} shows clock network routing, followed by signal routing and post-route optimization.
Redundant via insertion is performed in order to improve yield, and decap and standard cell fillers are inserted before streaming out GDS from the IC compiler.

Table \ref{tab:design_features} provides the details of clock tree QOR for the main clock. The clock tree was built in the slow process corner for around $18k$ sinks and achieved  skew of $240 ps$ with $2ns$ latency.
In Table \ref{tab:redundant_vias}, we present the percentage of redundant vias for various via layers.
We were able to achieve more than 98\% conversion of single to multi-cut vias for the lower via layers V1, V2, V3, V4, yet a lower percentage was achieved for higher layers.
In Table \ref{tab:pnr}, we present design statistics over various stages in the place \& route.
We remark that the standard cell count increases as the design moves from initial to final routing stages, primarily due to buffers/inverters inserted in the design to fix design rule violations, clock tree synthesis, and timing issues.
Our design started with 100\% HVT cells and ended up with 13.4\%, as HVT cells were swapped with RVT and LVT cells to address timing and DRV fixes (Table \ref{tab:pnr}).


\subsection{Sign-Off Analysis}
The design GDS obtained from the IC compiler is merged with the standard cell, memory, and IO pad GDS.
The seal ring is added on top, followed by the metal fill insertion procedure, 
and the corresponding GDS is generated, which is then merged with the design GDS to obtain the final GDS.
The sign-off DRC/LVS is then performed on this GDS using Cadence PVS.
The design parasitic extraction is done using STARRC-XT and the resulting SPEF along with the design netlist is fed to Synopsys Primetime-SI for static timing analysis.
The final DRC/LVS and timing clean GDS is then sent for fabrication.

\subsection{PLL design}
\begin{figure}[t!]
      \subfloat[Block diagram \label{fig:ADPLL_diagram}]{%
      \includegraphics[width=0.25\textwidth, height = 0.2\textwidth ]{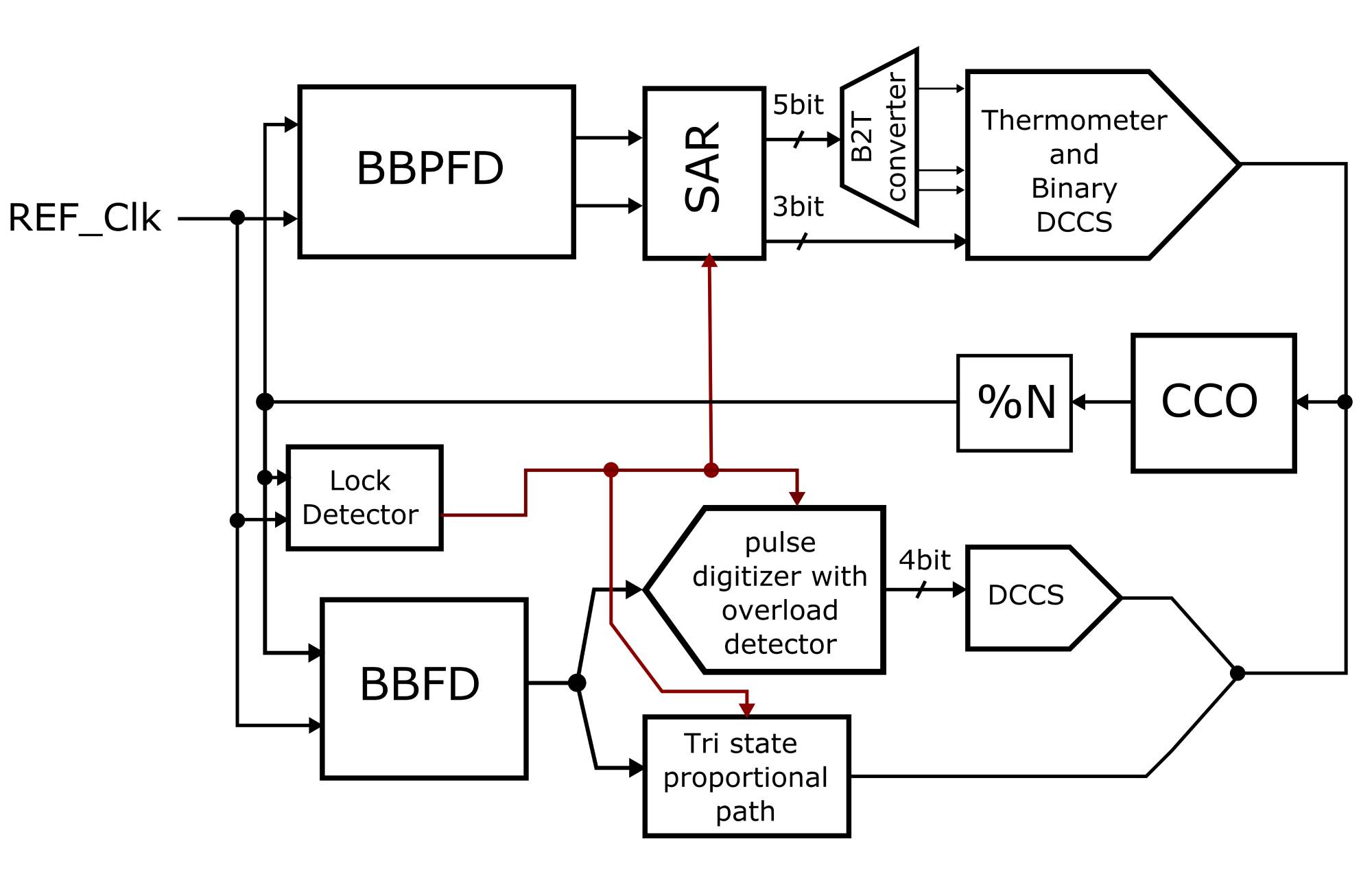}}
      \hspace{0.5mm}
      \subfloat[Layout in GF 55nm \label{fig:ADPLL_layout}]{%
      \includegraphics[width=0.23\textwidth, height = 0.2\textwidth ]{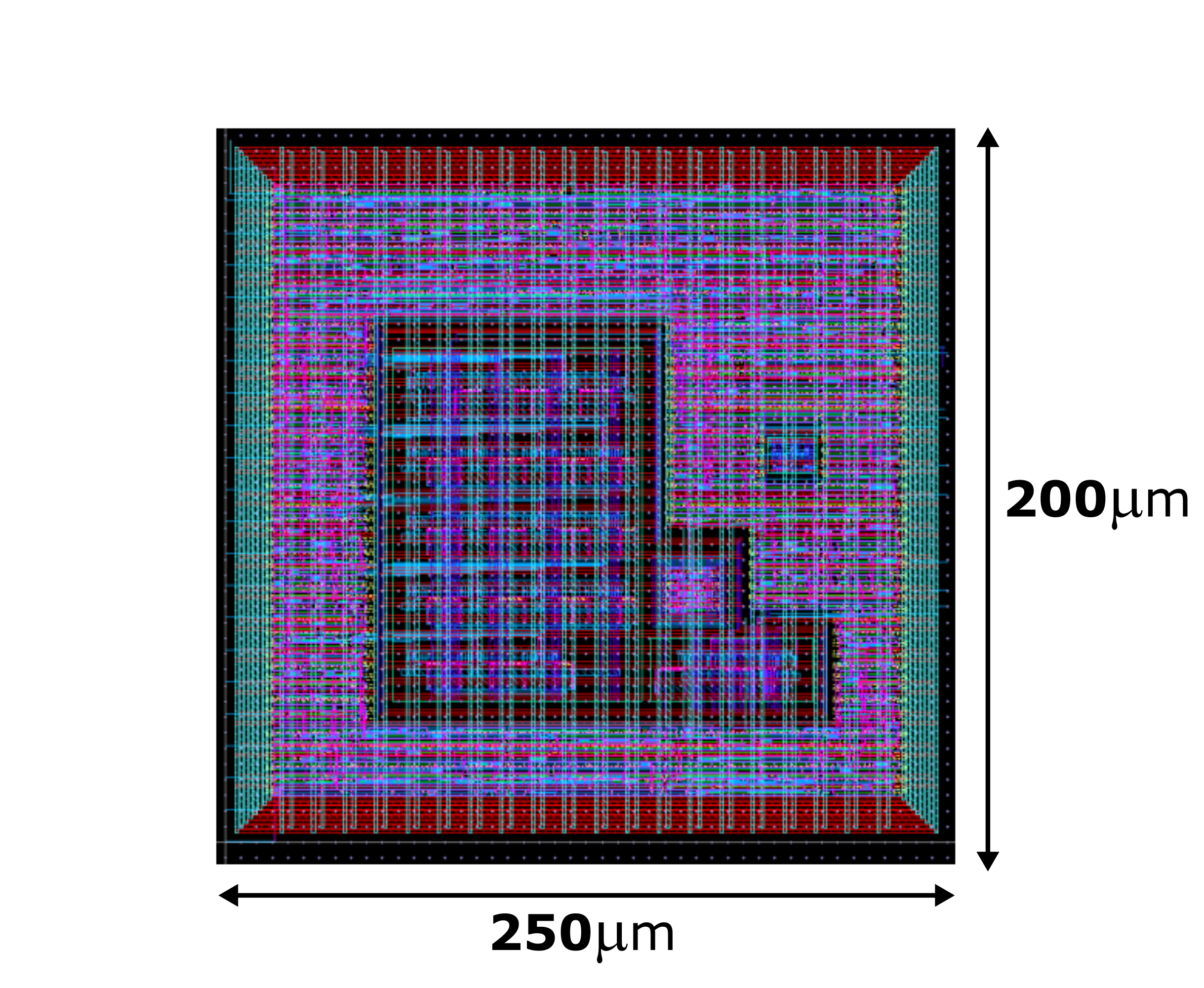}}\\
\\
      \subfloat[Power Straps \label{fig:powerstraps}]{%
      \includegraphics[width=0.25\textwidth, height = 0.2\textwidth ]{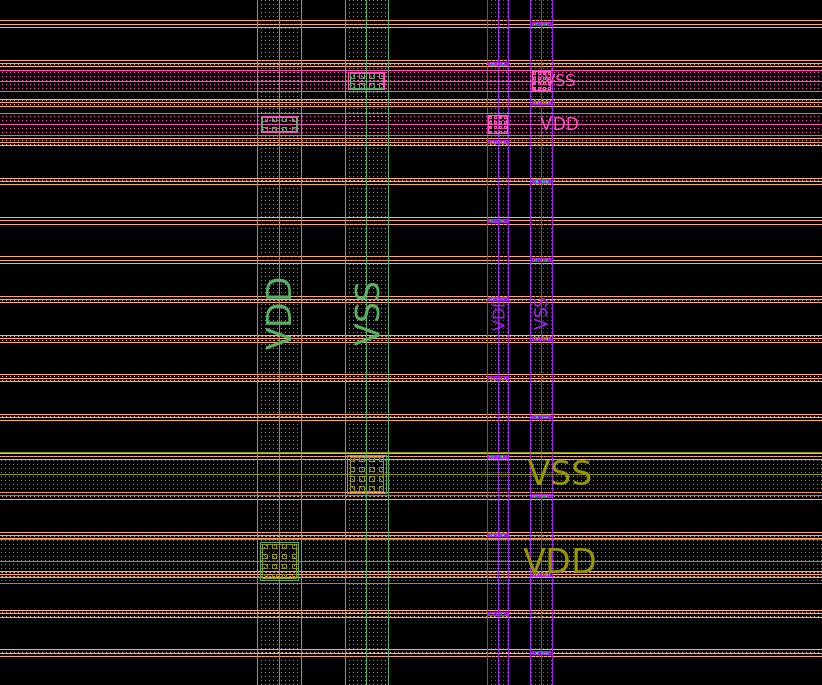}}
      \hspace{0.5mm}
      \subfloat[Power Ring\label{fig:powerring}]{%
      \includegraphics[width=0.23\textwidth, height = 0.2\textwidth ]{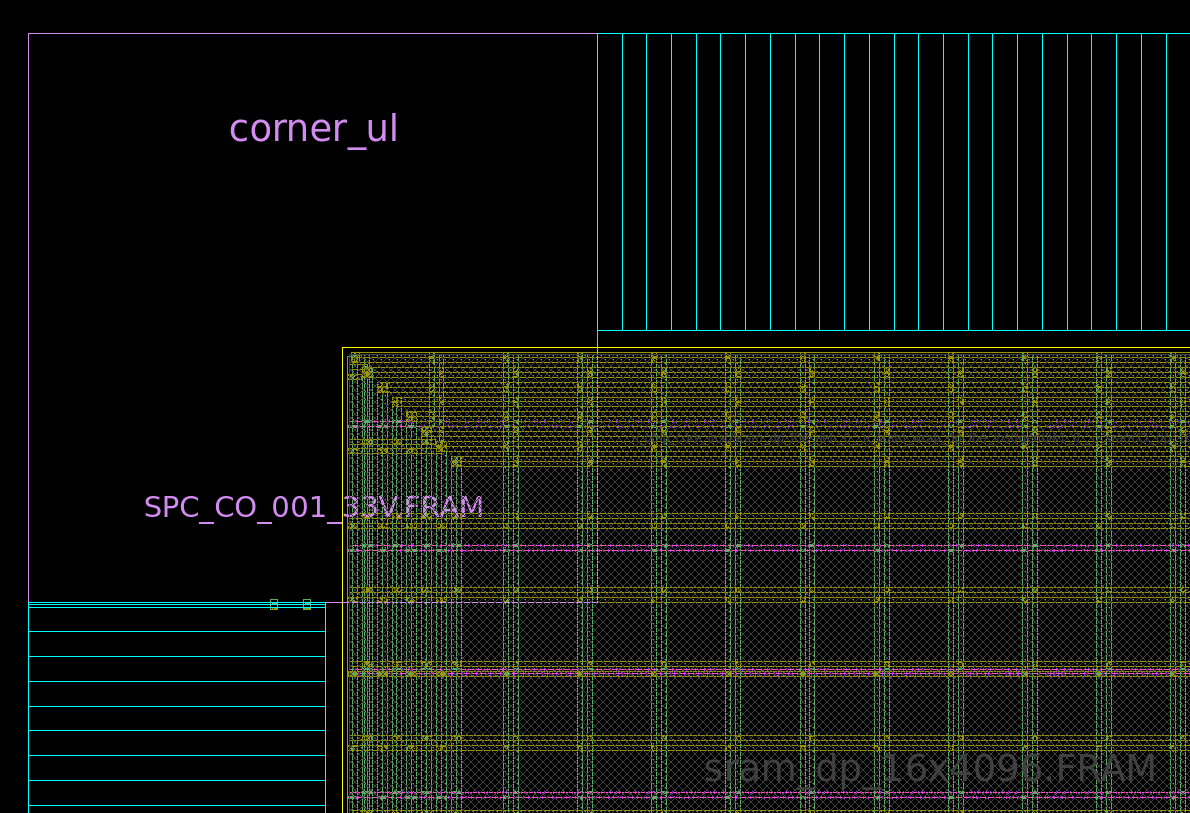}}\\
\caption{Proposed ADPLL design and additional Backend Results}
\end{figure}

In this section we discuss our efforts regarding the design of the PLL. Silicon area, power consumption, range of operation, and timing uncertainty are among the metrics that dictate the choices of a PLL architecture as well as implementation strategies.
For a given jitter performance, analog implementations require a loop filter with a big capacitor making the overall PLL size large.
On the other hand, a digital PLL provides an alternative with significantly smaller silicon area.
A digital implementation results in a low power implementation.
In addition, a wide range of operation is essential to run the chip at different frequencies.
This enables reusing the PLL in different designs, avoiding PLL redesign.
In this work, a compact, low power, and wide tuning range All Digital PLL (ADPLL) has been implemented.
The block diagram of the proposed ADPLL is shown in Fig. \ref{fig:ADPLL_diagram}, while the layout is presented in Fig. \ref{fig:ADPLL_layout}.
It is a dual-loop architecture with dedicated frequency and phase-locking loops.
The Frequency-Locking Loop (FLL) is a feedback loop which forces the frequency difference between the input signal and the oscillator output down to the capture range of the phase-locking loop.
Similarly, the Phase-Locking Loop (PLL) forces the phase error between input signal and the oscillator output to zero.

The capture range of the phase detector (PD) is a few percent of the reference clock frequency which is much smaller than the required tuning range of the ADPLL.
Consequently, a frequency-locking mechanism is required to pull the frequency of the local oscillator close to the operating frequency such that the PD with its narrow pull-in range can assume responsibility.
The choice of the frequency detector architecture used is based on the requirement that it should be easily interfaced with digital loop filters and allow a wide range of operation.
A digitized Phase and Frequency Detector (PFD) with a Successive Approximation Register (SAR) algorithm \cite{Rossi1996NonredundantSA}\cite{7452270}\cite{1084494} is employed to generate the appropriate digital control word to switch the required current value.

The phase-locking loop consists of a modified Alexander (Bang-Bang) phase detector \cite{Alexander1975ClockRF}\cite{Lee2004AnalysisAM}, and an all-digital loop filter in order to detect the phase error and generate an appropriate control signal.
The main requirements on the choice of the phase detector architecture originate from the need to use an all-digital loop filter, as it is compact, can easily be integrated with ASIC designs, and ported between technology nodes.
For these reasons, the phase detector should generate outputs which enable a digital implementation.
In this design, a modified version of the widely used Bang-Bang phase detector (BBPD) is proposed.
BBPD, also known as the early-late detection method, utilizes three data samples taken by three consecutive clock edges to determine whether a data transition is present or not, and if the clock leads or lags the data.
In the absence of data transitions, all three samples are equal, and no action is taken.

The loop filters in the FLL and PLL produce respectively digital control values from the instantaneous phase and frequency error signals generated by the frequency and phase detectors.
The outputs of the loop filters are used to allow the appropriate amount of supply current to the oscillator in order to correct for possible frequency or phase errors.
Since the oscillator frequency is controlled by current switching, segmented decoding is employed to avoid potential discontinuities and glitches.
This is achieved by implementing a combination of binary and unary weighted current sources.
To avoid any conflict between the frequency and phase correcting loops, a digital lock detector is used. 
We implement the ADPLL in GF 55nm CMOS technology.
It occupies an active area of $0.05mm^2$ and consumes $350 \uW$ from a supply of $1.1V$.

\subsection{Post-Silicon Validation}

\begin{figure}[!t]
	\centering
	\includegraphics[scale=0.09, trim={0 1in 0 0.6in},clip, width=0.4\textwidth,]{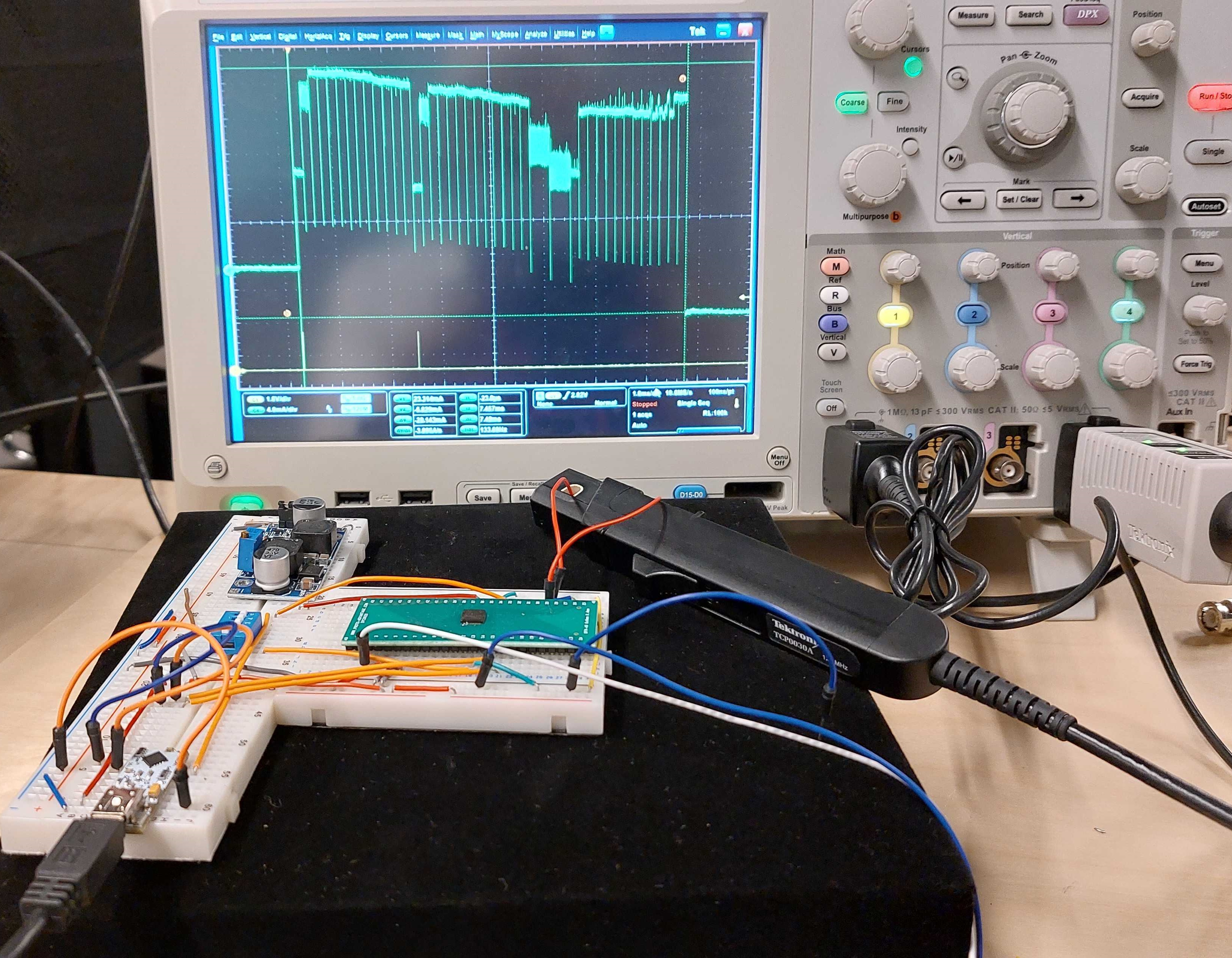}
	\caption{\cofhee{}'s validation and experimental setup} 
	\label{fig:setup}
\end{figure}

\cofhee{} is packaged in a 48-pin QFN, and connects to a breadboard via QFN to the DIP adapter for chip bring-up and testing.
For interfacing with a host computer, we use a UMFT230XA development board that features an FTDI chip for USB-to-UART conversion.
The UMFT230XA board provides a 3.3V supply for \cofhee{}'s IO pad, as well as a clock output which is used as reference clock.
Moreover, the required 1.2V supply was generated using a DC-DC adjustable step-down module that converts the 5V source of the UMFT230XA board.
In addition, a USB-to-UART breakout is used to receive the computation complete signal from \cofhee{}. For the power measurement, we use a Tektronix MSO 5204B oscilloscope with current probe. Our post-silicon validation setup is shown in Fig. \ref{fig:setup}, and confirms that the fabricated chip is fully functional. 

%% file: tables/pnr.tex
\begin{table}[t]
    \centering
    \vspace{0.45cm}
    \begin{tabular}{||l|r|r|r|r||}
        \hline
        {\bf Parameter}       & {\bf Initial} & {\bf Place} & {\bf CTS} & {\bf Route} \\ \hline \hline
        Standard cells        & 225,797       & 376,853     & 378,957   & 379,921     \\ \hline
        Sequential cells      & 18,686        & 18,686      & 18,686    & 18,686      \\ \hline
        Buffer/Inverter cells & 22,561        & 89,072      & 91,372    & 92,379      \\ \hline
        Std. Cell Utilization & 45\%          & 54\%        & 56.5\%    & 59\%        \\ \hline
        Signal nets           & 257,856       & 398,340     & 401,407   & 401,510     \\ \hline
        HVT cells             & 100\%         & 13.75\%     & 13.5\%    & 13.4\%      \\ \hline 
        RVT cells             & 0\%           & 17\%        & 12.1\%    & 12\%        \\ \hline
        LVT cells             & 0\%           & 69.25\%     & 74.4\%    & 74.6\%      \\ \hline
    \end{tabular}
    \caption{Design statistics through PnR}\label{tab:pnr}

    \vspace{-0.5cm}
\end{table}

        

%% file: figs/backend.tex

\begin{figure}[t!]
      \subfloat[Floorplan \label{fig:floorplan}]{%
      \includegraphics[width=0.15\textwidth, height = 0.15\textwidth ]{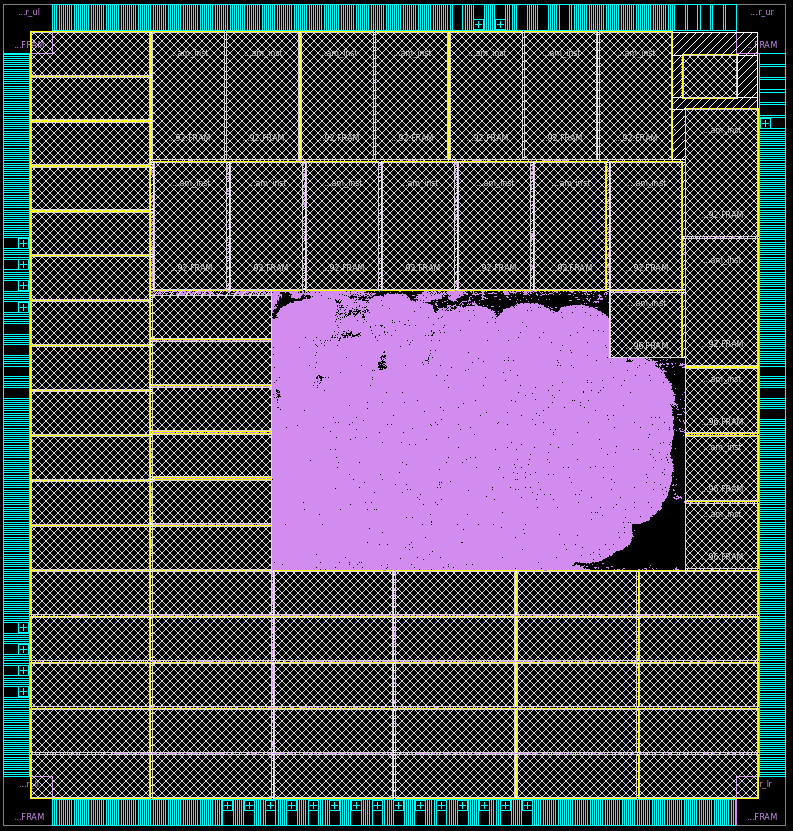}}
      \hspace{0.5mm}
      \subfloat[Power Network \label{fig:powernetwork}]{%
      \includegraphics[width=0.15\textwidth, height = 0.15\textwidth ]{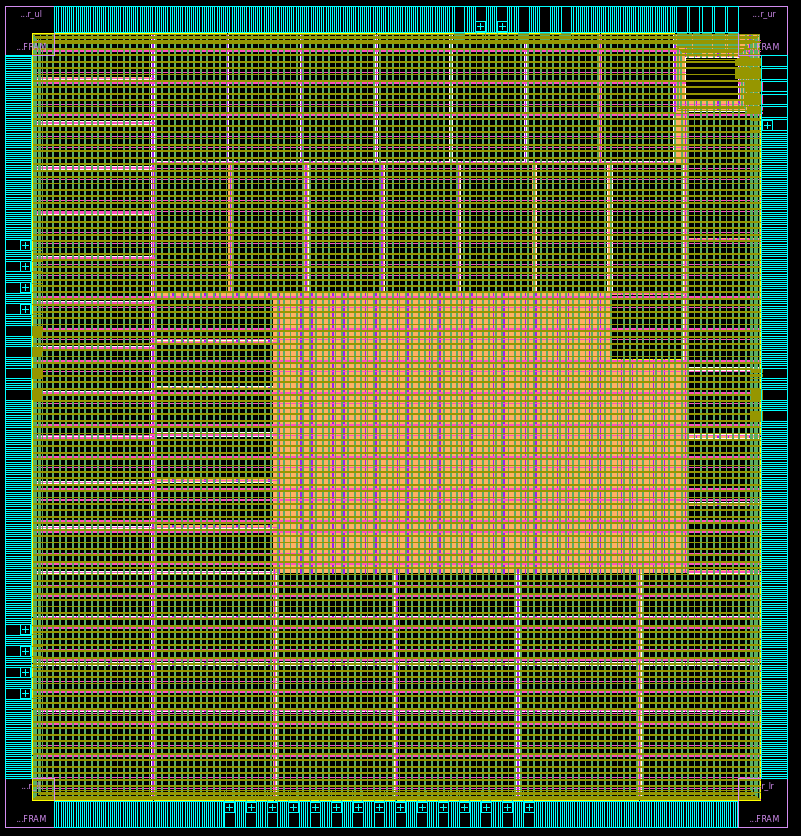}}
      \hspace{0.5mm}
      \subfloat[Clock Network\label{fig:clocknetwork}]{%
      \includegraphics[width=0.15\textwidth, height = 0.15\textwidth ]{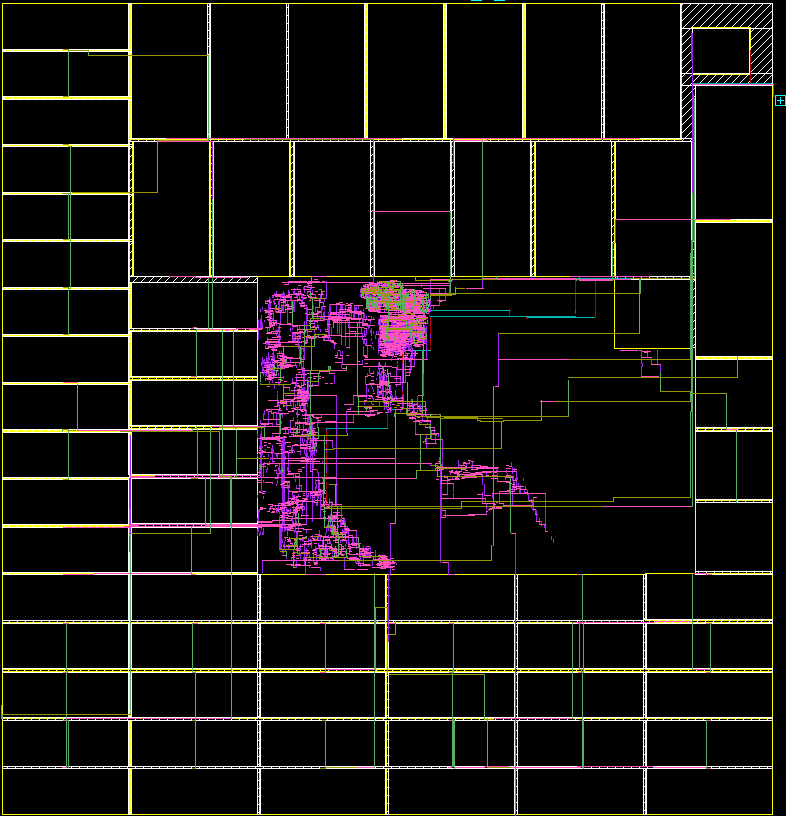}}\\
\\
      \subfloat[Power Pad \label{fig:powerpad}]{%
      \includegraphics[width=0.15\textwidth, height = 0.15\textwidth ]{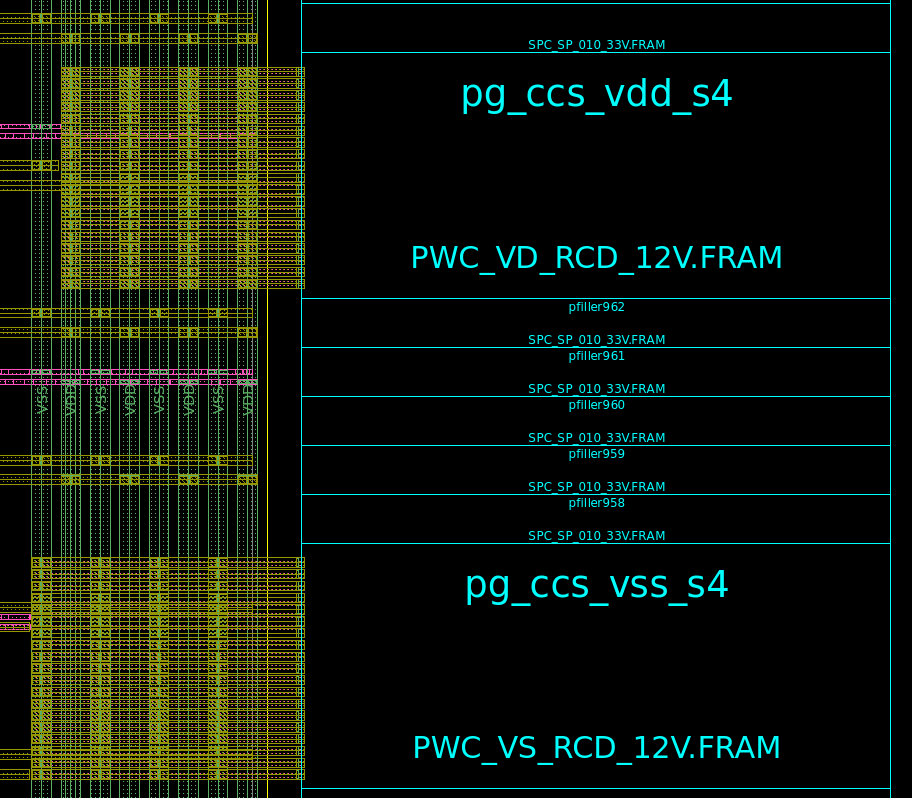}}
      \hspace{0.5mm}
      \subfloat[Power Connection\label{fig:powerconnection}]{%
      \includegraphics[width=0.15\textwidth, height = 0.15\textwidth ]{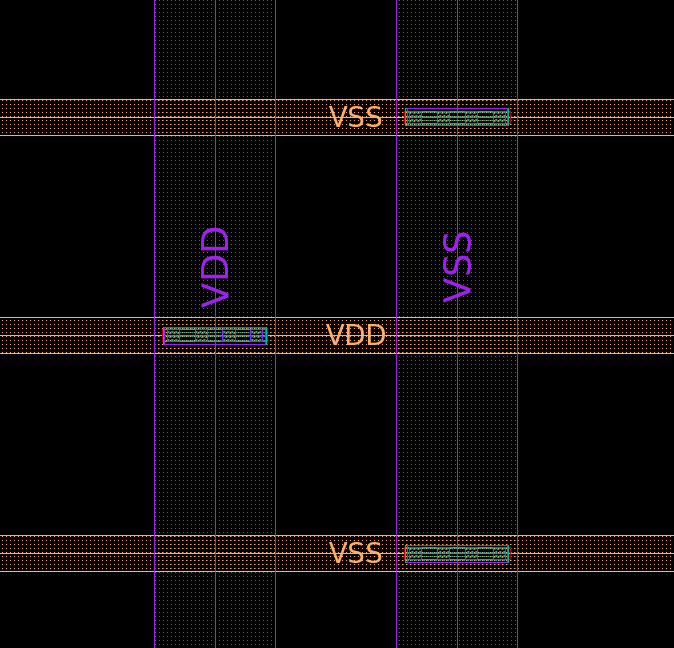}}
      \hspace{0.5mm}
      \subfloat[ Module placement\label{fig:placement}]{%
      \includegraphics[width=0.15\textwidth, height = 0.15\textwidth ]{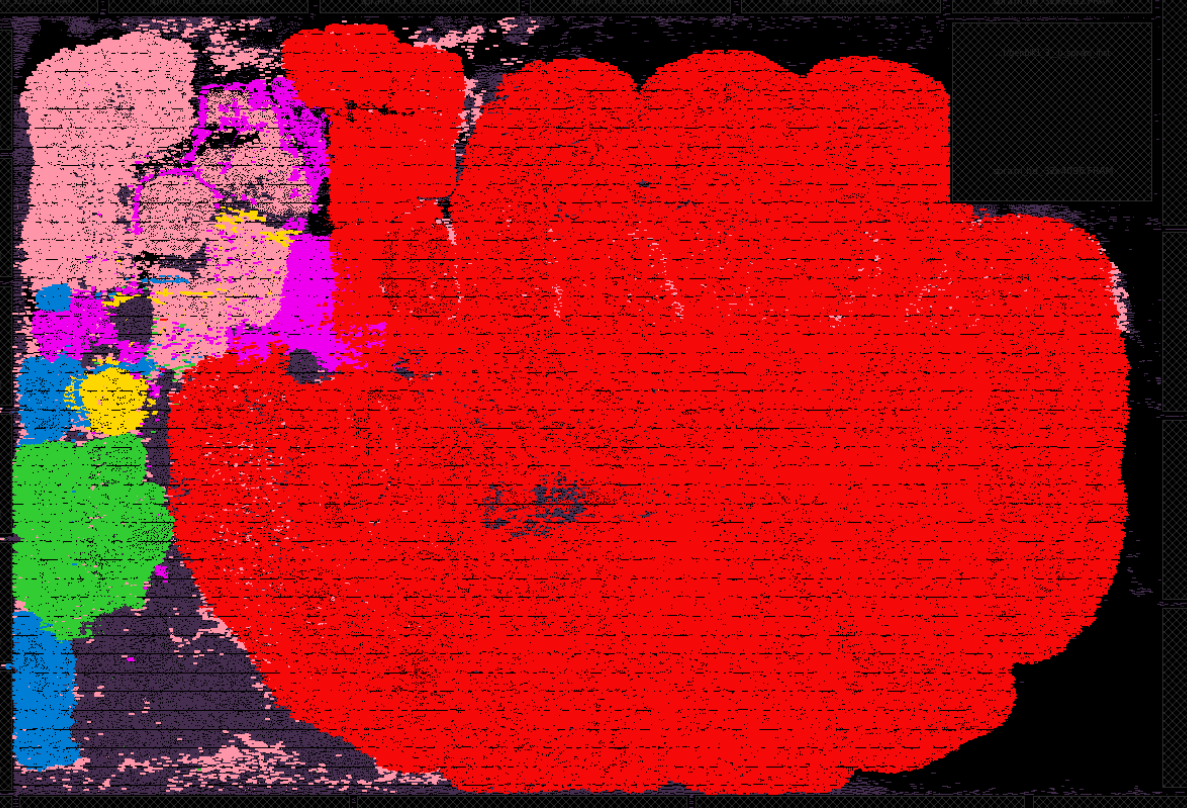}}\\
\caption{Backend Results}
\vspace{-0.2cm}
\vspace{-0.10in}
\end{figure}

%% file: tables/layout.tex
\begin{table}[H]
    \centering
    
    \begin{tabular}{||l|c||}
         \hline
         {\bf Parameter}         & {\bf Value}         \\ \hline \hline
        IU (Initial Utilization) & 45 \%               \\ \hline
        FU (Final Utilization)   & 59 \%               \\ \hline
        MA (Macro Area)          & 8,941,959 ${\um}^2$ \\ \hline
        HIO (IO PAD Height)      & 120 $\um$               \\ \hline
        CIO (Core to IO spacing) & 10 $\um$                \\ \hline
        A (Aspect ratio)         & 1.05                \\ \hline
        CA (std cell Area)       & 1,963,585 ${\um}^2$ \\ \hline
        CW (Core Width)          & 3400 $\um$              \\ \hline
        CH (Core Height)         & 3582 $\um$              \\ \hline
        DW (Die Width)           & 3660 $\um$              \\ \hline
        DH (Die Height)          & 3842 $\um$              \\ \hline
    \end{tabular}
    \caption{Layout physical parameters}\label{tab:layout}
\end{table}

%% file: tables/performance.tex
\begin{table}[H]
    \centering
    
    \begin{tabular}{||c|r|r|c|c||}
    \hline
    \multirow{2}{*}{Algorithm} & \multicolumn{2}{c|}{Latency} & \multicolumn{2}{c||}{Power {\footnotesize ($mW$)}} \\ 
    {}                         & \multicolumn{1}{c}{(cc)}     & \multicolumn{1}{c|}{($\mu$s)} & \multicolumn{1}{c}{avg.}   & \multicolumn{1}{c||}{peak} \\ \hline \hline
    \multicolumn{5}{||c||}{$n = 2^{12}$} \\ \hline
    PolyMul                    &  83,777 & 335.1 & 22.9 & 30.4 \\ 
    NTT                        &  24,841 &  99.4 & 24.5 & 30.4 \\ 
    iNTT                       &  29,468 & 117.9 & 19.9 & 27.2 \\ \hline \hline
    \multicolumn{5}{||c||}{$n = 2^{13}$} \\ \hline
    PolyMul                    & 179,045 & 716.2 & 21.2 & 29.7 \\ 
    NTT                        &  53,535 & 214.1 & 24.4 & 29.7 \\ 
    iNTT                       &  62,770 & 251.1 & 18.3 & 23.9 \\ \hline

    \end{tabular}
    \caption{\cofhee{} performance for $n=\{2^{12},2^{13}\}$}
    \label{tab:sys_latency}
\end{table}

%% file: tables/eda_tools.tex
\begin{table}[H]
    \centering
    \begin{tabular}{||l|c||}
        \hline
        \multicolumn{1}{||c|}{\textbf{Stage}} & \multicolumn{1}{c||}{\textbf{Tool}} \\ \hline \hline
        \multirow{2}*{Place and Route}        & Synopsys                            \\
        {}                                    & IC compiler                         \\ \hline
        Interconnect                     & Synopsys                            \\
        parasitic extraction                      & STAR-RCXT                           \\ \hline
        Static timing                     & Synopsys                            \\
        analysis                                 & Prime-Time-SI                       \\ \hline
        GDS merging and                       & Cadence                             \\
        layout modification                   & Virtuoso                            \\ \hline
        Physical                     & Cadence PVS                         \\
        verification                                  & System                              \\ \hline
    \end{tabular}
    \caption{Stages and EDA tools.}
    \label{tab:stages_tools}
\end{table}


%% file: tables/redundant_vias.tex
\begin{table}[H]
    \centering
    
    \begin{tabular}{||c|c|c|c||}
        \hline
        \multirow{2}*{\bf Layer} & {\bf multi-}                  & {\bf total}         & {\bf multi-}                   \\
        {}                       & {\bf cut {\footnotesize(\#)}} & {\footnotesize(\#)} & {\bf cut {\footnotesize(\%)}}  \\ \hline \hline
        V1 & 21,659 & 21,945 & 98.70 \\ \hline
        V2 & 21,732 & 21,844 & 99.49 \\ \hline
        V3 & 21,991 & 22,035 & 99.80 \\ \hline
        V4 & 26,391 & 26,455 & 99.76 \\ \hline
        WT &  2,438 &  2,450 & 99.51 \\ \hline
        WA &  1,390 &  1,393 & 99.78 \\ \hline
    \end{tabular}
    \caption{Redundant via statistics}\label{tab:redundant_vias}
\end{table}


%% file: 4_evaluation.tex
\section{Experimental Evaluation}
\label{sec:eval}

\subsection{Power and latency}
We have measured the power consumption and latency for several operations supported by \cofhee{} using polynomial degrees of $n=\{2^{12}, 2^{13}\}$. The results of these measurements are summarized in Table \ref{tab:sys_latency}. Our measurements show that the NTT operation results in the highest peak power consumption, while the Hadamard and Inverse NTT (iNTT) operations consume less power. The iNTT operation includes a multiplication with a constant ($n^{-1}$) and a decimation in frequency operation, which results in a lower average power consumption compared to NTT, due to the lower power consumption of the constant multiplication. However, iNTT takes more cycles to execute than NTT. To conclude, \cofhee{} requires a power supply with a peak power rating of around $30mA$ and an average power of around $25mA$ to perform polynomial multiplication in a fraction of a millisecond.


\subsection{Comparison to CPU}
\begin{figure}[t!]
      \subfloat[Time for all towers \label{fig:times}]{%
      \includegraphics[width=0.98\linewidth]{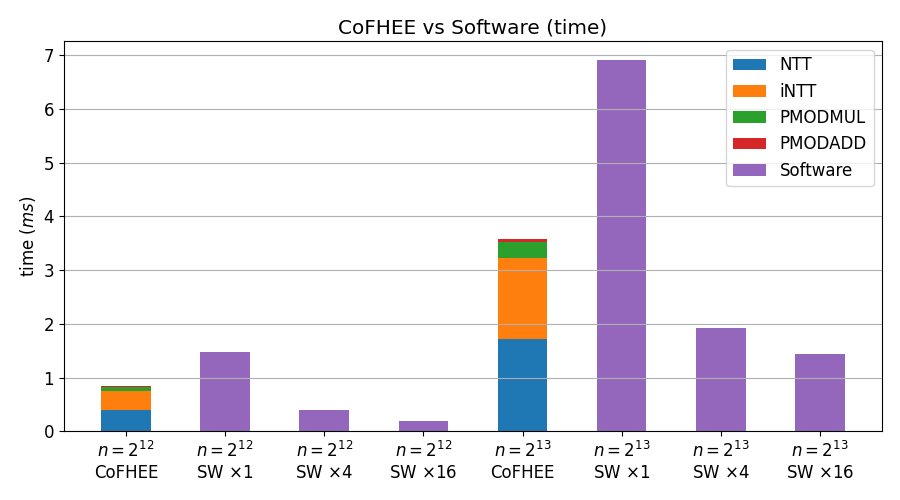}}
      \hspace{0.5mm}
      \subfloat[Power \label{fig:power}]{%
      \includegraphics[width=0.98\linewidth]{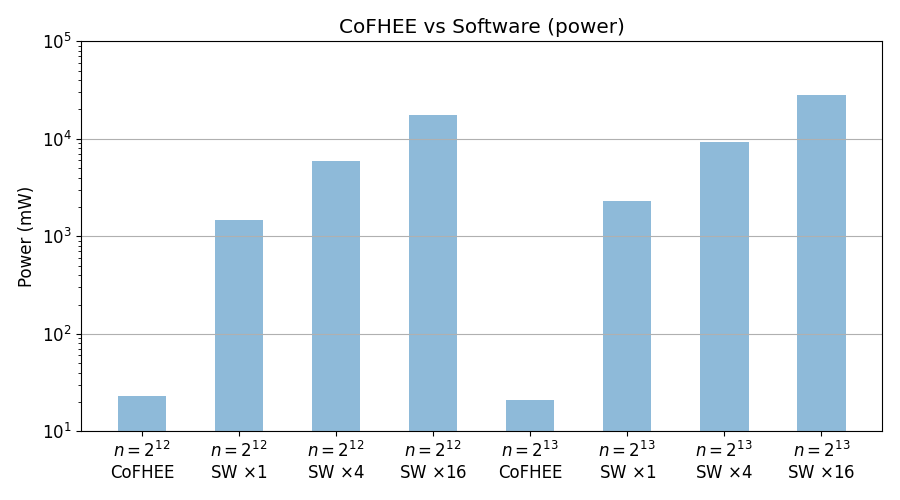}}
\caption{Comparison to CPU execution}
\label{fig:results}
\end{figure}

We have performed a comparative analysis of \cofhee{} against a software implementation in terms of both execution time and power consumption by executing a ciphertext multiplication without relinearization using one instance of \cofhee{}. For the software implementation, we have employed the Microsoft SEAL 3.7 library \cite{seal}, executing it on an AMD Ryzen 7 5800h (TSMC 7nm FinFET) processor operating at 3.8Ghz, paired with 16GB of RAM on Ubuntu 20.04 LTS, and we have gathered power measurements using \texttt{powertop}. We have established the polynomial degree as $n = \{2^{12}, 2^{13}\}$ and $\log q = \{109, 218\}$ bits, which provide a security level of 128 bits against classical computers. The results of this comparison are illustrated in Fig. \ref{fig:results}.


For $(n, \log q) = (2^{12}, 109)$, we break SEAL's 109-bit modulus into two smaller moduli of 54 and 55 bits using RNS for faster computation in the native 64-bit architecture.
Each of these two towers must perform the ciphertext multiplication according to Eq. \ref{eq:ctmul}.
For the same operation, \cofhee{} natively supports 128 bits, therefore, it requires only one tower.
As shown in Fig. \ref{fig:times}, the software implementation takes $1.5ms$ to operate on the two towers running on a single thread, while \cofhee{} needs $0.84ms$ to finalize the ciphertext multiplication.
Regarding power (Fig. \ref{fig:power}), \cofhee{} is two orders of magnitude more efficient as it requires $22mW$ of power, while the software implementation uses $1.48W$.

\begin{figure}[t]
    \begin{minipage}[t][][t]{0.20\textwidth}
        \input{tables/area}
    \end{minipage}
    \hspace{-0.08\textwidth}
    \begin{minipage}[t][][t]{0.5\textwidth}
        \input{tables/design_features}
    \end{minipage}
\end{figure}

When $(n, \log q) = (2^{13}, 218)$, we set the software implementation to use four towers of around 55 bits ($54 + 54 + 55 + 55 = 218$) and \cofhee{} requires two towers ($109 + 109 = 218$).
For the four $<64$-bit towers, SEAL spends $6.91 ms$ when running on one thread, while \cofhee{} takes $3.58ms$ to operate on the two $<128$-bit towers.
Power readings are $21.2mW$ for \cofhee{} and $2.3W$ for the software implementation.
We also evaluate SEAL using multi-thread ($4 \times$ for four and $16 \times$ for sixteen threads in Fig. \ref{fig:results}). As expected, the execution time reduces as we add threads to the point of becoming faster than a single instance of \cofhee{}. We also notice diminishing returns as we add extra threads. It is important to consider that CoFHEE contains only one PE because it is manufactured with an area footprint of $12mm^2$ in 55nm technology, which is smaller compared to the CPU we analyzed. The CPU has an area of $180mm^2$ and is built using the 7nm technology node. If CoFHEE could accommodate four PEs (which would allow it to perform radix-4 butterfly operations in a pipeline), its performance would increase by a factor of $=\sim 4$ (The NTT cycle count is calculated as $(N/radix) * log_{base-radix}(N))$, and the area would increase by only $1.9mm^2$ for the addition of three additional PEs (reference: Table VIII). This improvement exceeds the performance achieved with 16 threads.

Nevertheless, the power consumption increases near linearly. Thus, the software Power-Delay Product (PDP) is lowest when using a single thread, which gets a PDP equal to $2.22 \ W \cdot ms$ ($15.9 \ W \cdot ms$), while \cofhee{}'s PDP is $18.5 \cdot 10^{-3} \ W \cdot ms$ ($75.9 \cdot 10^{-3} \ W \cdot ms$) for $n=2^{12}$ ($n=2^{13}$), 2 to 3 orders of magnitude more efficient.
While at its current instantiation \cofhee{} provides a modest speed-up compared to a modern CPU, it is much smaller in size and uses a much bigger technology node. Therefore, when fabricated in the same node and normalized for the area used, \cofhee{} will offer orders of magnitude improvement over CPU execution of FHE.

\subsection{End-to-End Applications}

We evaluated \cofhee{} based on its performance in two comprehensive applications. The first is CryptoNets \cite{dowlin2016cryptonets}, a method that leverages neural networks and Fully Homomorphic Encryption for efficient secure inference. The second is a test against the logistic regression inference as detailed in \cite{sarkar2023privacy}. The expected processing times for both of these applications when using \cofhee{} are provided in Table \ref{tab:e2e}.

In our analysis, the execution runtime was assessed in relation to the number of operations involved in the application. Specifically, operations such as ciphertext-ciphertext multiplications ($ct \cdot ct$) and additions ($ct+ct$), as well as ciphertext-plaintext multiplications ($ct \cdot pt$), were considered. For CryptoNets, the total operations consisted of 457,550 ct-ct additions, 449,000 ct-pt multiplications, and 10,200 ct-ct multiplications. This subsequently necessitated 10,200 relinearization operations. In contrast, the logistic regression inference entailed 168,298 ct-ct additions, 49,500 ct-pt multiplications, and 128,700 combined ct-ct multiplications and relinearizations.

Upon evaluating these operations using \cofhee{}, the execution times were found to be 88.35 seconds for CryptoNets and 377.6 seconds for Logistic Regression. In comparison to CPU execution, \cofhee{} exhibited speedups of 2.23x for CryptoNets and 1.46x for logistic regression.





\input{tables/e2e_table}

%% file: tables/area.tex
%

 \begin{table}[H]
     \centering
     \begin{tabular}{||l|r|r||}
         \hline
          \multirow{2}*{\bf Module} & \multicolumn{1}{c|}{\bf Area} & \multicolumn{1}{c||}{\bf Delay}  \\
          {}                        & \multicolumn{1}{c|}{($mm^2$)} & \multicolumn{1}{c||}{\bf $(ns)$} \\ \hline \hline
          3 DP SRAMs                & 5.3506                        & 4.22 \\ \hline
          4 SP SRAMs                & 3.2036                        & 4.19 \\ \hline
          PE                        & 0.6394                        & 5.65 \\ \hline
          CM0 SRAM                  & 0.4062                        & 6.13 \\ \hline
          AHB                       & 0.0747                        & 5.76 \\ \hline
          GPCFG                     & 0.0534                        & 7.03 \\ \hline
          ARM CM0                   & 0.0354                        & 5.24 \\ \hline
          MDMC                      & 0.0273                        & 4.16 \\ \hline
          SPI                       & 0.0202                        & 7.74 \\ \hline
          DMA                       & 0.0075                        & 7.17 \\ \hline
          UART                      & 0.0065                        & 5.66 \\ \hline
          GPIO                      & 0.0035                        & 6.73 \\ \hline
          Others                    & 0.0063                        & -    \\ \hline \hline
          {\bf Total}               & {\bf 9.8345}                  & - \\ \hline
     \end{tabular}
     \vspace{-1em}
     \captionsetup{justification=centering}
    \captionsetup{width=1.3\linewidth}
     \caption{Part estimations}
     \label{tab:area}
 \end{table}


%% file: tables/design_features.tex
\begin{table}[H]
    \centering
    \begin{tabular}{||l|r||}
        \hline
        \textbf{Parameter}    & \textbf{Value} \\ \hline \hline
        Width                 & 3660 $\um$           \\ \hline
        Height                & 3842 $\um$           \\ \hline
        Signal pads           &   26           \\ \hline
        PG pads               &   11           \\ \hline
        PLL bias pads         &    8           \\ \hline
        Memories              &   68           \\ \hline
        clock name            & HCLK           \\ \hline
        CTS synth. corner     & slow           \\ \hline
        Levels                & 26             \\ \hline
        Sinks                 & 18413          \\ \hline
        Clock tree buffers    & 464            \\ \hline
        Global Skew           & 240 $ps$       \\ \hline
        Longest Ins. delay    & 2.079 $ns$     \\ \hline
        Shortest Ins. delay   & 1.838 $ns$     \\ \hline
    \end{tabular}
    \caption{Design statistics}
    \label{tab:design_features}
\end{table}

%% file: tables/e2e_table.tex
\begin{table}[]
\centering
\caption{End-to-End Application comparison}
 \label{tab:e2e}
\begin{tabular}{l|lll}
                    & CPU & CoFHEE & Speedup \\ \hline
CryptoNets \cite{dowlin2016cryptonets}          &  $197 s$   &    $88.35 s$    & $2.23\times$    \\
Logistic Regression \cite{sarkar2023privacy} &   $550.25 s$  &   $377.6 s$     & $1.46\times$    \\ \hline
\end{tabular}
\end{table}


%% file: 5_relatedWork.tex
\section{Related Work}
\label{sec:related_work}

\input{tables/f1_heax_comp}
FHE acceleration has become a major research interest in the recent years.
So far, many different approaches have been proposed, consequently leading to a state-of-the-art that is comprised of FHE acceleration solutions that are based on software, GPU, FPGA, and ASIC \cite{jung2020heaan}, \cite{HEAX}, \cite{DesignMert20}, \cite{hsu2020vlsi}, \cite{reagen2021cheetah}, \cite{cilardo2016securing}, \cite{cousins2016designing}, \cite{bradbury2021ntt}.
Each method has demonstrated considerable advantages, however due to hardware differences, establishing an accurate comparison point between each technology is not feasible, therefore we focus on ASIC designs in our discussion. Table \ref{tab:F1_HEAX_comp} summarizes existing works on ASIC and FPGA.

In addition, there are also works that have focused on the acceleration of polynomial multiplications using ASIC designs, however they do not focus in FHE applications.
While some may focus on different variants of homomorphic encryption such as PHE and SHE, many works focus purely on developing an acceleration method without taking into account the application the method can potentially be applied to \cite{asif18} \cite{boateng18} \cite{ding2017modular} \cite{asif16} \cite{kuang16} \cite{wang12} \cite{sakiyama11} \cite{knezevic10}.

CuFHE is an FHE library designed for use with CUDA-enabled GPUs. It uses the TFHE scheme and introduces a new method for handling FHE computations. However, its approach takes about $ 4 \mu\text{s} $  to process NTT operations with polynomial sizes of $2^{14}$, which is slower than what CoFHEE offers.

The closest work to \cofhee{} is F1, a work proposed by Feldmann et al. that focuses on developing an accelerator for the execution of FHE programs \cite{feldmann2021f1}.
The solution can be described as a wide-vector processor with functional units specialized to FHE primitives like modular arithmetic, NTT, and structured permutations.
However, there is limited information about the backend process of the design, and whether the proposed architecture can be placed on real silicon without routing congestion, as well as power and clock issues. The API is not clear either, since the focus is more on architecture exploration and less on getting a realized chip.

When comparing \cofhee{} with F1, we need to consider the differences between the area, frequency, technology nodes, as well as the input size (bits) of each design. Consequently, to conduct a fair and accurate evaluation, we have normalized the performance in terms of the area and the scaling factor between the technology nodes. To obtain the scaling factor, we synthesized the Barrett modular multiplier using the GF7nm technology library, which is the same as that used for F1. The results indicate that the scaling factor reduces the area by $16.7\times$ and the critical path by $3.7\times$.
We compare the NTT operation of CoFHEE and F1 in terms of time per area with a polynomial degree of $8$k and a modulus size of $128$ bits. 
We compare the PE and RF area in F1 against the PE and MDMC area of CoFHEE. We exclude memory as F1 contains extra on-chip memory to support higher-level operations, such as key switching. Also, we consider that F1 has to do RNS to split 128-bit coefficients into 32-bit towers.
Upon normalization of the performance metrics, it was found that F1 executed $7.21\cdot{10}^{-5}$ NTT operations per $ns$ per $mm^2$, while \cofhee{} achieved $4.54 \cdot {10}^{-4}$, resulting in a speedup of $6.3\times$. This enhancement in performance is mainly attributed to the use of a pipelined Barrett multiplier, as opposed to an iterative Montgomery multiplier.



CraterLake \cite{samardzic2022craterlake}, ARK \cite{kim2022ark}, and BTS \cite{kim2022bts} also represent three hardware accelerators dedicated to enhancing the performance of Fully Homomorphic Encryption (FHE) computations. While each is proficient at handling comprehensive FHE tasks on-chip and showcases notable efficiency enhancements, their physical footprints are extensive. Specifically, they cover areas of $472.3 mm^2$, $418.3 mm^2$, and $373.6 mm^2$, respectively. Such expansive dimensions pose challenges for fabrication as the wafer would necessitate even more space during the production phase. This factor renders them impractical for manufacturing and, subsequently, for potential real-world utilization as opposed to \cofhee{} that only occupies $12mm^2$ of area. Nevertheless, after conducting the same comparison process as F1, \cofhee{} exhibits $1.39\times$, $46.19\times$, and $4.72\times$ improvement against CraterLake, BTS, and ARK respectively when compared on the NTT operation with a polynomial degree of $n=2^{13}$.


Besides ASIC, there is a number of works that propose FPGA-based solutions for FHE acceleration. HEAX \cite{riazi_heax} is a parallelizable architecture focusing on number-theoretic transform (NTT), and Roy~\cite{roy2019fpga} presents an FPGA-based co-processor for FHE computation using the BFV scheme. A comparison of the NTT performance against the related work is presented in Table \ref{tab:F1_HEAX_comp}, in terms of clock cycles. The performance per $m m^2$ efficiency metric cannot be accurately calculated, however, as we cannot map FPGA resources to silicon area. In terms of raw numbers, F1 reports a $1,733\times$ improvement over HEAX, so we expect \cofhee{} to also be significantly more efficient than HEAX.






%% file: tables/f1_heax_comp.tex
\begin{table*}[t]
    \centering
    \resizebox{2\columnwidth}{!}{%
    \begin{tabular}{||c|c|c|c|c|c|c|c|c|c||}
        \hline
         \multirow{2}*{\bf Design} & {\bf Technology} & {\bf Max} & {\bf log q} & {\bf Area} & {\bf Power } &   {\bf Freq.} & {\bf Clock } & {\bf Efficiency*} & {\bf Silicon} \\
                                {}  &       &    {\bf n}     &    ($bits$) &   ($mm^2$ | LUT/FF/BRAM/DSP) & {\bf ($W$)}       &    ($Mhz$) &   {\bf Cycles*} &   {\bf (Performance per $mm^2$)}       &   {\bf Proven}  \\ \hline \hline
           \cofhee{}                &         ASIC - GF 55$nm$ &$2^{14}$ &         128 &         12 & $2.3 \cdot 10^{-2}$                 &        250 &   53248       &    $4.54$ $\cdot$ $10^{-4}$           & \ding{108}   \\ \hline
           F1 \cite{feldmann2021f1} &      ASIC - GF 14/12$nm$ &  $2^{14}$ &         32  &      151.4 & $1.8 \cdot 10^2$               &        1000 & 476       &       $7.21$ $\cdot$ $10^{-5}$         & \ding{109}   \\ \hline
           CraterLake \cite{samardzic2022craterlake} &      ASIC - 14/12$nm$ & $2^{16}$  &         28  &      472.3 & $3.2 \cdot 10^2$               &        1000 &   22     &      $3.26$ $\cdot$ $10^{-4}$          & \ding{109}   \\ \hline
           BTS \cite{kim2022bts} &      ASIC - 7$nm$ &  $2^{17}$ &         64  &      373.6 & $1.6 \cdot 10^2$               &        1200 &   554     &       $9.83$ $\cdot$ $10^{-6}$        & \ding{109}   \\ \hline
           ARK \cite{kim2022ark} &      ASIC - 7$nm$ & $2^{16}$  &         64  &      418.3 & $2.8 \cdot 10^2$               &        1000 &    104    &       $9.62$ $\cdot$ $10^{-5}$         & \ding{109}   \\ \hline
           HEAX \cite{riazi_heax}   &       FPGA - Intel Arria10 GX 1150  & $2^{14}$ &         27  &         $582148$ / $1554005$ / $3986$ / $2018$ &   -                  &       300 &  1536       &     $\dagger$           & N/A   \\ \hline
           Roy \cite{roy2019fpga}   &        Xilinx Zynq UltraScale+ ZCU102  &  $2^{12}$ &         30  &          $63522$ / $25622$ / $400$ / $200$ &   -                  &       200 & 16425       &      $\dagger$         & N/A    \\ \hline
    \end{tabular}
    }
    \caption{Comparative table and performance of the NTT operation against related work}
    \footnotesize{$\dagger$ No information is available to accurately map FPGA resources to silicon area }\\
    \footnotesize{$*$ Evaluation of Efficiency and clock cycles are performed for $n=2^{13}$ }\\
    \label{tab:F1_HEAX_comp}
\end{table*}

%% file: 6_discussion.tex
\section{Discussion}
\label{sec:discussion}

\subsection{\cofhee{} scalability}

To support operations on higher polynomial degrees $n \ge 2^{14}$ with $II = 1$, \cofhee{} needs more area for memories, which increase linearly to the polynomial degree.
As the memory size increases, memory read latency increases, which leads to a minor reduction in clock frequency.
However, if the goal is to improve performance for a particular $n$, there are a few options:
First, we could perform more than one one butterfly operations per cycle.
For that, one has to duplicate the multiplier pool and add more memories to increase parallel accesses of operands.
For NTT with $II=1$, two dual-port memories and one multiplier pool are required.
Doubling this improves throughput by close to $2 \times$ as one can split the polynomial of degree $n$ into two smaller polynomials of degree $n/2$, perform NTT on these smaller polynomials in parallel, recombine them and perform the last stage of NTT of the original polynomial.
This effectively means that $\log (n) - 1$ stages operate with $II=0.5$, while the last stage has an $II$ of $1$. Furthermore, a second approach is to increase the number of memories and processing units available.
The former increases the number of polynomials on chip, enabling more complex scheduling methods which reduce communication costs with the host.
With more memories and processing units, parallel operations on non-dependent data is also possible.
Combined better scheduling and parallel processing, doubling \cofhee{}'s resources would more than double the throughput.

\cofhee{}'s AHB interface connects to scalable and flexible components.
It supports up to 16 controllers, memories, and peripherals and, with ARM Cortex M0, it has 28 bits for addressing internal memory.
Hence, it is possible to increase the total memory size from 1 MB (currently used) to 256 MB.
For faster communication, UART or SPI can be replaced with protocols such as Peripheral Component Interconnect Express.

\subsection{Lessons Learned}

Using the AHB bus to fetch operands and load them to the computation unit avoids complex MUX structures \cite{riazi_heax_2020} that come from non trivial addressing of operands and twiddle factors in every NTT stage.
During NTT, \cofhee{}'s address generation unit generates the relevant addresses for operands and twiddle factors at every cycle and issues a bus transaction.

Although the addition of dual-port memories improves the throughput significantly, their area is $2 \times$ the area of single-port memories of the same size.
Therefore, it is important to minimize the number of dual-port memories without affecting  throughput.
In \cofhee{}, the number of dual-port memories is three.
Thus, two of them can be used for computation, while the third is used along with DMA to buffer the next polynomial for execution.

While separate twiddle factors have been used for NTT and iNTT in previous works \cite{riazi_heax_2020}, \cofhee{} uses the same twiddle factors for both operations by combining MDMC and DMA operations.
Thus, it reduces the required chip area for FHE acceleration.


%% file: 7_conclusion.tex
\section{conclusion}
\label{sec:conc}
In this work, we presented a year-long effort to design, implement, and validate \cofhee{}, a co-processor for processing of low-level polynomial operations targeting Fully Homomorphic Encryption execution.
\cofhee{} is a $12mm^2$ design fabricated in $55nm$ technology node and it natively supports polynomial degrees of up to $2^{14}$, as well as coefficient sizes up to 128 bits.
The architecture design is highly scalable and can be  expanded to support higher polynomial degrees, more polynomials on chip, and more processing units given a larger design area. This paper presents all required steps for a fully functional silicon, from the RTL design to fabrication and validation. The RTL code and functional units of \cofhee{} are open-sourced, to serve as components for future FHE acceleration work.

\section{Acknowledgments}
The work was funded by the Center for Cyber Security Abu Dhabi (CCSAD) at New York University Abu Dhabi and also the Global PhD Fellowship program. 

\section{Resources}
All developed methodologies, code, and experiments in this
work are open-sourced, and can be found under the following link: \url{https://github.com/momalab/CoFHEE}

%% file: main.bbl
\begin{thebibliography}{10}
\providecommand{\url}[1]{#1}
\csname url@samestyle\endcsname
\providecommand{\newblock}{\relax}
\providecommand{\bibinfo}[2]{#2}
\providecommand{\BIBentrySTDinterwordspacing}{\spaceskip=0pt\relax}
\providecommand{\BIBentryALTinterwordstretchfactor}{4}
\providecommand{\BIBentryALTinterwordspacing}{\spaceskip=\fontdimen2\font plus
\BIBentryALTinterwordstretchfactor\fontdimen3\font minus \fontdimen4\font\relax}
\providecommand{\BIBforeignlanguage}[2]{{%
\expandafter\ifx\csname l@#1\endcsname\relax
\typeout{** WARNING: IEEEtran.bst: No hyphenation pattern has been}%
\typeout{** loaded for the language `#1'. Using the pattern for}%
\typeout{** the default language instead.}%
\else
\language=\csname l@#1\endcsname
\fi
#2}}
\providecommand{\BIBdecl}{\relax}
\BIBdecl

\bibitem{barron2013cloud}
C.~Barron, H.~Yu, and J.~Zhan, ``Cloud computing security case studies and research,'' in \emph{World Congress on Engineering}, 2013.

\bibitem{gentry2009thesis}
C.~Gentry, \emph{A fully homomorphic encryption scheme}.\hskip 1em plus 0.5em minus 0.4em\relax Stanford University, 2009.

\bibitem{dsoni}
D.~Soni, M.~Nabeel, H.~Gamil, O.~Mazonka, B.~Reagen, R.~Karri, and M.~Maniatakos, ``Design space exploration of modular multipliers for asic fhe accelerators,'' in \emph{2023 24th International Symposium on Quality Electronic Design (ISQED)}.\hskip 1em plus 0.5em minus 0.4em\relax IEEE, 2023, pp. 1--8.

\bibitem{zhang}
N.~Zhang, H.~Gamil, P.~Brinich, B.~Reynwar, A.~Al~Badawi, N.~Neda, D.~Soni, K.~Canida, Y.~Polyakov, P.~Broderick \emph{et~al.}, ``Towards full-stack acceleration for fully homomorphic encryption,'' \emph{IEEE HPEC}, 2022.

\bibitem{soni2023rpu}
D.~Soni, N.~Neda, N.~Zhang, B.~Reynwar, H.~Gamil, B.~Heyman, M.~Nabeel, A.~Al~Badawi, Y.~Polyakov, K.~Canida \emph{et~al.}, ``Rpu: The ring processing unit,'' in \emph{2023 IEEE International Symposium on Performance Analysis of Systems and Software (ISPASS)}.\hskip 1em plus 0.5em minus 0.4em\relax IEEE, 2023, pp. 272--282.

\bibitem{dai2014accelerating}
W.~Dai, Y.~Dor{\"o}z, and B.~Sunar, ``Accelerating \uppercase{NTRU} based homomorphic encryption using \uppercase{GPU}s,'' in \emph{HPEC}, 2014.

\bibitem{cousins2012update}
D.~B. Cousins, K.~Rohloff, C.~Peikert, and R.~Schantz, ``An update on \uppercase{SIPHER} (scalable implementation of primitives for homomorphic encryption)—\uppercase{FPGA} implementation using simulink,'' in \emph{2012 IEEE Conference on High Performance Extreme Computing}, 2012.

\bibitem{gentry2011implementing}
C.~Gentry and S.~Halevi, ``Implementing gentry’s fully-homomorphic encryption scheme,'' in \emph{Annual international conference on the theory and applications of cryptographic techniques}, 2011.

\bibitem{doroz2014accelerating}
Y.~Dor{\"o}z, E.~{\"O}zt{\"u}rk, and B.~Sunar, ``Accelerating fully homomorphic encryption in hardware,'' \emph{IEEE Transactions on Computers}, vol.~64, no.~6, pp. 1509--1521, 2014.

\bibitem{feldmann2021f1}
A.~Feldmann, N.~Samardzic, A.~Krastev, S.~Devadas, R.~Dreslinski, K.~Eldefrawy, N.~Genise, C.~Peikert, and D.~Sanchez, ``F1: A fast and programmable accelerator for fully homomorphic encryption (extended version),'' \emph{arXiv preprint arXiv:2109.05371}, 2021.

\bibitem{bfv}
J.~Fan and F.~Vercauteren, ``Somewhat practical fully homomorphic encryption,'' \emph{IACR Cryptology ePrint Archive}, vol. 2012, p. 144, 2012.

\bibitem{bgv}
Z.~Brakerski, C.~Gentry, and V.~Vaikuntanathan, ``(leveled) fully homomorphic encryption without bootstrapping,'' \emph{ACM Transactions on Computation Theory (TOCT)}, vol.~6, no.~3, pp. 1--36, 2014.

\bibitem{cggi}
I.~Chillotti, N.~Gama, M.~Georgieva, and M.~Izabach{\`e}ne, ``Faster fully homomorphic encryption: Bootstrapping in less than 0.1 seconds,'' in \emph{Advances in Cryptology -- ASIACRYPT 2016}, 2016.

\bibitem{ckks}
J.~H. Cheon, A.~Kim, M.~Kim, and Y.~Song, ``Homomorphic encryption for arithmetic of approximate numbers,'' in \emph{ASIACRYPT}, 2017.

\bibitem{rlwe}
V.~Lyubashevsky, C.~Peikert, and O.~Regev, ``On ideal lattices and learning with errors over rings,'' \emph{J. ACM}, vol.~60, no.~6, 2013.

\bibitem{alchemy}
\BIBentryALTinterwordspacing
E.~Crockett, C.~Peikert, and C.~Sharp, ``Alchemy: A language and compiler for homomorphic encryption made easy,'' in \emph{Proceedings of the 2018 ACM SIGSAC Conference on Computer and Communications Security}, ser. CCS '18.\hskip 1em plus 0.5em minus 0.4em\relax New York, NY, USA: Association for Computing Machinery, 2018, p. 1020–1037. [Online]. Available: \url{https://doi.org/10.1145/3243734.3243828}
\BIBentrySTDinterwordspacing

\bibitem{cingulata}
\BIBentryALTinterwordspacing
S.~Carpov, P.~Dubrulle, and R.~Sirdey, ``Armadillo: A compilation chain for privacy preserving applications,'' in \emph{Proceedings of the 3rd International Workshop on Security in Cloud Computing}, ser. SCC '15.\hskip 1em plus 0.5em minus 0.4em\relax New York, NY, USA: Association for Computing Machinery, 2015, p. 13–19. [Online]. Available: \url{https://doi.org/10.1145/2732516.2732520}
\BIBentrySTDinterwordspacing

\bibitem{e3}
E.~Chielle, O.~Mazonka, N.~G. Tsoutsos, and M.~Maniatakos, ``E$^3$: A framework for compiling c++ programs with encrypted operands,'' Cryptology ePrint Archive, Report 2018/1013, 2018, online: \url{https://eprint.iacr.org/2018/1013}, GitHub repository: \url{https://github.com/momalab/e3}.

\bibitem{ngraph}
\BIBentryALTinterwordspacing
F.~Boemer, A.~Costache, R.~Cammarota, and C.~Wierzynski, ``Ngraph-he2: A high-throughput framework for neural network inference on encrypted data,'' in \emph{Proceedings of the 7th ACM Workshop on Encrypted Computing \& Applied Homomorphic Cryptography}, ser. WAHC'19.\hskip 1em plus 0.5em minus 0.4em\relax New York, NY, USA: Association for Computing Machinery, 2019, p. 45–56. [Online]. Available: \url{https://doi.org/10.1145/3338469.3358944}
\BIBentrySTDinterwordspacing

\bibitem{palisade}
Y.~Polyakov, K.~Rohloff, G.~W. Ryan, and D.~Cousins, ``{PALISADE} lattice cryptography library user manual (v.1.10.6),'' 2020, \url{https://palisade-crypto.org/documentation}.

\bibitem{ramparts}
\BIBentryALTinterwordspacing
D.~W. Archer, J.~M. Calder\'{o}n~Trilla, J.~Dagit, A.~Malozemoff, Y.~Polyakov, K.~Rohloff, and G.~Ryan, ``Ramparts: A programmer-friendly system for building homomorphic encryption applications,'' in \emph{Proceedings of the 7th ACM Workshop on Encrypted Computing \& Applied Homomorphic Cryptography}, ser. WAHC'19.\hskip 1em plus 0.5em minus 0.4em\relax New York, NY, USA: Association for Computing Machinery, 2019, p. 57–68. [Online]. Available: \url{https://doi.org/10.1145/3338469.3358945}
\BIBentrySTDinterwordspacing

\bibitem{seal}
``{M}icrosoft {SEAL} (release 3.7),'' \url{https://github.com/Microsoft/SEAL}, Sep. 2021, microsoft Research, Redmond, WA.

\bibitem{bridging}
\BIBentryALTinterwordspacing
E.~Chielle, O.~Mazonka, H.~Gamil, and M.~Maniatakos, ``Accelerating fully homomorphic encryption by bridging modular and bit-level arithmetic,'' in \emph{Proceedings of the 41st IEEE/ACM International Conference on Computer-Aided Design}, ser. ICCAD '22.\hskip 1em plus 0.5em minus 0.4em\relax New York, NY, USA: Association for Computing Machinery, 2022. [Online]. Available: \url{https://doi.org/10.1145/3508352.3549415}
\BIBentrySTDinterwordspacing

\bibitem{hestd}
M.~Albrecht, M.~Chase, H.~Chen, J.~Ding, S.~Goldwasser, S.~Gorbunov, S.~Halevi, J.~Hoffstein, K.~Laine, K.~Lauter, S.~Lokam, D.~Micciancio, D.~Moody, T.~Morrison, A.~Sahai, and V.~Vaikuntanathan, ``Homomorphic encryption security standard,'' HomomorphicEncryption.org, Toronto, Canada, Tech. Rep., November 2018.

\bibitem{cooley-tukey}
J.~W. Cooley and J.~W. Tukey, ``An algorithm for the machine calculation of complex fourier series,'' \emph{Mathematics of Computation}, vol.~19, no.~90, pp. 297--301, 1965.

\bibitem{gentleman-sande}
W.~M. Gentleman and G.~Sande, ``Fast fourier transforms: For fun and profit,'' in \emph{Proceedings of the November 7-10, 1966, Fall Joint Computer Conference}, 1966, p. 563–578.

\bibitem{kim2015private}
M.~Kim and K.~Lauter, ``Private genome analysis through homomorphic encryption,'' \emph{BMC medical informatics and decision making}, 2015.

\bibitem{pmt}
E.~Chielle, H.~Gamil, and M.~Maniatakos, ``Real-time private membership test using homomorphic encryption,'' in \emph{2021 Design, Automation Test in Europe Conference Exhibition (DATE)}, 2021, pp. 1282--1287.

\bibitem{gursoy}
G.~Gürsoy, E.~Chielle, C.~M. Brannon, M.~Maniatakos, and M.~Gerstein, ``Privacy-preserving genotype imputation with fully homomorphic encryption,'' \emph{Cell Systems}, vol.~13, no.~2, pp. 173--182.e3, 2022.

\bibitem{modmulalgocomp}
A.~Bosselaers, R.~Govaerts, and J.~Vandewalle, ``Comparison of three modular reduction functions,'' in \emph{Annual International Cryptology Conference}, 1993, pp. 175--186.

\bibitem{noauthor_amba_2001}
ARM, ``{AMBA} 3 {AHB}-{Lite} {Protocol} {Specification},'' p.~72, 2001.

\bibitem{cryptohandbook}
A.~J. Menezes, P.~C. Van~Oorschot, and S.~A. Vanstone, \emph{Handbook of applied cryptography}.\hskip 1em plus 0.5em minus 0.4em\relax CRC press, 2018.

\bibitem{Rossi1996NonredundantSA}
A.~Rossi and G.~Fucili, ``Nonredundant successive approximation register for a/d converters,'' \emph{Electronics Letters}, vol.~32, pp. 1055--1057, 1996.

\bibitem{7452270}
B.~Murmann, ``The successive approximation register adc: a versatile building block for ultra-low- power to ultra-high-speed applications,'' \emph{IEEE Communications Magazine}, vol.~54, no.~4, pp. 78--83, 2016.

\bibitem{1084494}
H.~Russell, ``An improved successive-approximation register design for use in a/d converters,'' \emph{IEEE Transactions on Circuits and Systems}, vol.~25, no.~7, pp. 550--554, 1978.

\bibitem{Alexander1975ClockRF}
J.~D. Alexander, ``Clock recovery from random binary signals,'' \emph{Electronics Letters}, vol.~11, pp. 541--542, 1975.

\bibitem{Lee2004AnalysisAM}
J.~Lee, K.~S. Kundert, and B.~Razavi, ``Analysis and modeling of bang-bang clock and data recovery circuits,'' \emph{IEEE Journal of Solid-State Circuits}, vol.~39, pp. 1571--1580, 2004.

\bibitem{dowlin2016cryptonets}
R.~Gilad-Bachrach, N.~Dowlin, K.~Laine, K.~Lauter, M.~Naehrig, and J.~Wernsing, ``Cryptonets: Applying neural networks to encrypted data with high throughput and accuracy,'' in \emph{International conference on machine learning}.\hskip 1em plus 0.5em minus 0.4em\relax PMLR, 2016, pp. 201--210.

\bibitem{sarkar2023privacy}
E.~Sarkar, E.~Chielle, G.~Gursoy, L.~Chen, M.~Gerstein, and M.~Maniatakos, ``Privacy-preserving cancer type prediction with homomorphic encryption,'' \emph{Scientific reports}, vol.~13, no.~1, p. 1661, 2023.

\bibitem{samardzic2022craterlake}
N.~Samardzic, A.~Feldmann, A.~Krastev, N.~Manohar, N.~Genise, S.~Devadas, K.~Eldefrawy, C.~Peikert, and D.~Sanchez, ``Craterlake: a hardware accelerator for efficient unbounded computation on encrypted data,'' in \emph{Proceedings of the 49th Annual International Symposium on Computer Architecture}, 2022, pp. 173--187.

\bibitem{kim2022bts}
S.~Kim, J.~Kim, M.~J. Kim, W.~Jung, J.~Kim, M.~Rhu, and J.~H. Ahn, ``Bts: An accelerator for bootstrappable fully homomorphic encryption,'' in \emph{Proceedings of the 49th Annual International Symposium on Computer Architecture}, 2022, pp. 711--725.

\bibitem{kim2022ark}
J.~Kim, G.~Lee, S.~Kim, G.~Sohn, M.~Rhu, J.~Kim, and J.~H. Ahn, ``Ark: Fully homomorphic encryption accelerator with runtime data generation and inter-operation key reuse,'' in \emph{2022 55th IEEE/ACM International Symposium on Microarchitecture (MICRO)}.\hskip 1em plus 0.5em minus 0.4em\relax IEEE, 2022, pp. 1237--1254.

\bibitem{riazi_heax}
M.~S. Riazi, K.~Laine, B.~Pelton, and W.~Dai, ``\uppercase{HEAX}: An architecture for computing on encrypted data,'' in \emph{ASPLOS '20}, 2020, p. 1295–1309.

\bibitem{roy2019fpga}
S.~S. Roy, F.~Turan, K.~Jarvinen, F.~Vercauteren, and I.~Verbauwhede, ``\uppercase{FPGA}-based high-performance parallel architecture for homomorphic computing on encrypted data,'' in \emph{HPCA}, 2019.

\bibitem{jung2020heaan}
W.~Jung, E.~Lee, S.~Kim, K.~Lee, N.~Kim, C.~Min, J.~H. Cheon, and J.~H. Ahn, ``Heaan demystified: Accelerating fully homomorphic encryption through architecture-centric analysis and optimization,'' \emph{arXiv preprint arXiv:2003.04510}, 2020.

\bibitem{HEAX}
\BIBentryALTinterwordspacing
M.~S. Riazi, K.~Laine, B.~Pelton, and W.~Dai, ``Heax: An architecture for computing on encrypted data,'' in \emph{Proceedings of the Twenty-Fifth International Conference on Architectural Support for Programming Languages and Operating Systems}, ser. ASPLOS '20.\hskip 1em plus 0.5em minus 0.4em\relax New York, NY, USA: Association for Computing Machinery, 2020, p. 1295–1309. [Online]. Available: \url{https://doi.org/10.1145/3373376.3378523}
\BIBentrySTDinterwordspacing

\bibitem{DesignMert20}
A.~C. Mert, E.~{\"O}zt{\"u}rk, and E.~Sava{\c{s}}, ``Design and implementation of encryption/decryption architectures for bfv homomorphic encryption scheme,'' \emph{IEEE Transactions on Very Large Scale Integration (VLSI) Systems}, vol.~28, no.~2, pp. 353--362, 2019.

\bibitem{hsu2020vlsi}
H.-J. Hsu and M.-D. Shieh, ``Vlsi architecture of polynomial multiplication for bgv fully homomorphic encryption,'' in \emph{2020 IEEE International Symposium on Circuits and Systems (ISCAS)}.\hskip 1em plus 0.5em minus 0.4em\relax IEEE, 2020, pp. 1--4.

\bibitem{reagen2021cheetah}
B.~Reagen, W.-S. Choi, Y.~Ko, V.~T. Lee, H.-H.~S. Lee, G.-Y. Wei, and D.~Brooks, ``Cheetah: Optimizing and accelerating homomorphic encryption for private inference,'' in \emph{2021 IEEE International Symposium on High-Performance Computer Architecture (HPCA)}.\hskip 1em plus 0.5em minus 0.4em\relax IEEE, 2021, pp. 26--39.

\bibitem{cilardo2016securing}
A.~Cilardo and D.~Argenziano, ``Securing the cloud with reconfigurable computing: An fpga accelerator for homomorphic encryption,'' in \emph{2016 Design, Automation \& Test in Europe Conference \& Exhibition (DATE)}.\hskip 1em plus 0.5em minus 0.4em\relax IEEE, 2016, pp. 1622--1627.

\bibitem{cousins2016designing}
D.~B. Cousins, K.~Rohloff, and D.~Sumorok, ``Designing an fpga-accelerated homomorphic encryption co-processor,'' \emph{IEEE Transactions on Emerging Topics in Computing}, vol.~5, no.~2, pp. 193--206, 2016.

\bibitem{bradbury2021ntt}
J.~Bradbury, N.~Drucker, and M.~Hillenbrand, ``Ntt software optimization using an extended harvey butterfly,'' \emph{Cryptology ePrint Archive}, 2021.

\bibitem{asif18}
S.~{Asif}, O.~{Andersson}, J.~{Rodrigues}, and Y.~{Kong}, ``65-nm cmos low-energy rns modular multiplier for elliptic-curve cryptography,'' \emph{IET Computers Digital Techniques}, vol.~12, no.~2, pp. 62--67, 2018.

\bibitem{boateng18}
R.~Boateng~Nti and K.~Ryoo, ``Asic design of low area rsa crypto-core based on montgomery multiplier,'' \emph{International Journal of Engineering and Technology}, vol.~7, pp. 278--283, 08 2018.

\bibitem{ding2017modular}
J.~Ding and S.~Li, ``A modular multiplier implemented with truncated multiplication,'' \emph{IEEE Transactions on Circuits and Systems II: Express Briefs}, vol.~65, no.~11, pp. 1713--1717, 2017.

\bibitem{asif16}
S.~Asif and Y.~Kong, ``Highly parallel modular multiplier for elliptic curve cryptography in residue number system,'' \emph{Circuits, Systems, and Signal Processing}, vol.~36, 05 2016.

\bibitem{kuang16}
S.~{Kuang}, K.~{Wu}, and R.~{Lu}, ``Low-cost high-performance vlsi architecture for montgomery modular multiplication,'' \emph{IEEE Transactions on Very Large Scale Integration (VLSI) Systems}, vol.~24, no.~2, pp. 434--443, 2016.

\bibitem{wang12}
S.~{Wang}, W.~{Lin}, J.~{Ye}, and M.~{Shieh}, ``Fast scalable radix-4 montgomery modular multiplier,'' in \emph{2012 IEEE International Symposium on Circuits and Systems (ISCAS)}, 2012, pp. 3049--3052.

\bibitem{sakiyama11}
K.~Sakiyama, M.~Knežević, J.~Fan, B.~Preneel, and I.~Verbauwhede, ``Tripartite modular multiplication,'' \emph{Integration}, vol.~44, pp. 259--269, 09 2011.

\bibitem{knezevic10}
M.~{Knezevic}, F.~{Vercauteren}, and I.~{Verbauwhede}, ``Faster interleaved modular multiplication based on barrett and montgomery reduction methods,'' \emph{IEEE Transactions on Computers}, vol.~59, no.~12, pp. 1715--1721, 2010.

\bibitem{riazi_heax_2020}
\BIBentryALTinterwordspacing
M.~S. Riazi, K.~Laine, B.~Pelton, and W.~Dai, ``{HEAX}: {An} {Architecture} for {Computing} on {Encrypted} {Data},'' \emph{arXiv:1909.09731 [cs]}, Jan. 2020, arXiv: 1909.09731. [Online]. Available: \url{http://arxiv.org/abs/1909.09731}
\BIBentrySTDinterwordspacing

\end{thebibliography}
